\documentclass{article}
\usepackage{graphicx} % Required for inserting images
\usepackage{xspace}
\newcommand{\netmob}{\texttt{NetMob25}\xspace}
\usepackage{enumitem}
\usepackage{xcolor}
\usepackage{tabularx}     
\usepackage{booktabs}     
\usepackage{subcaption}
\usepackage{hyperref}
\usepackage{tikz}
\usepackage{float}
\usetikzlibrary{arrows, positioning}
\usepackage{booktabs}
\usepackage{geometry}
\usepackage{amsmath}
\usepackage{caption}
\usepackage{tabularx}
\usepackage{subcaption}
\usetikzlibrary{arrows.meta, positioning}
\usepackage{graphicx}
\usepackage{array}

\title{The NetMob25 Dataset: A High-resolution Multi-layered View of Individual Mobility in Greater Paris Region}

\author{
  Alexandre Chasse\textsuperscript{1}\thanks{These authors contributed equally to this work.},
  Anne J. Kouam\textsuperscript{2}\footnotemark[1],
  Aline C. Viana\textsuperscript{2}\footnotemark[1],
  Razvan Stanica\textsuperscript{2,3}\footnotemark[1], \\
  Wellington V. Lobato\textsuperscript{2},
  Geymerson Ramos\textsuperscript{2},
  Geoffrey Deperle\textsuperscript{2}, \\
  Abdelmounaim Bouroudi\textsuperscript{2}, 
  Suzanne Bussod\textsuperscript{1},
  Fernando Molano\textsuperscript{2} \\ \vspace{0.2cm}
  \textsuperscript{1}IFPEN, France \quad
  \textsuperscript{2}Inria, France \quad
  \textsuperscript{3}INSA Lyon, France
}

\date{June 2025}

\begin{document}

\maketitle

\begin{abstract}

High-quality mobility data remains scarce despite growing interest from researchers and urban stakeholders in understanding individual-level movement patterns. The \netmob Data Challenge addresses this gap by releasing
an unique GPS-based mobility dataset derived from the \textit{EMG 2023 GNSS-based mobility survey} conducted in the Île-de-France region (Greater Paris area), France. This \netmob dataset captures detailed daily mobility over a full week for 3,337 volunteer residents aged 16 to 80, collected between mid-October 2022 and mid-May 2023. Each participant was equipped with a dedicated GPS tracking device, configured to record location points every 2–3 seconds, and asked to maintain a digital (or paper) logbook of their trips. All inferred mobility traces were algorithmically processed and subsequently validated through follow-up phone interviews with the participants.

The dataset is organized into three complementary databases: (i) an Individuals database describing each participant’s demographic, socioeconomic, and household characteristics; (ii) a Trips database containing over 80,000 annotated displacements with metadata such as departure and arrival times, transport modes (including multimodal combinations), and trip purposes; and (iii) a Raw GPS Traces database comprising approximately 500 million high-frequency points. In addition, a statistical weighting mechanism is provided to infer population-level estimates from the sample. Prior to the release of the \netmob dataset, an extensive anonymization pipeline was applied to the GPS traces to ensure compliance with GDPR regulations while preserving the analytical utility of the data. In addition, \netmob dataset will be made available to participants only upon acceptance of the \netmob Data Challenge’s \textit{Terms and Conditions} and the signing of a Non-Disclosure Agreement (NDA).

This paper presents the survey design, the data collection protocol, the cleaning methodology, and characterizes the released dataset across key dimensions. The \netmob dataset offers a valuable new resource for mobility research, supporting studies on travel behavior, transport mode choice, activity patterns, teleworking, and more. 
%\ali{The \netmob dataset will be made available to participants after acceptance of the challenge’s Terms and Conditions and NDA signature, fostering methodological innovation and advancing research addressing societal and urban mobility challenges.}
%\ali{To foster methodological innovation and advancing research addressing societal and urban mobility challenges, the \netmob dataset will be made available to participants upon acceptance of the challenge’s \textit{Terms and Conditions} and the signing of a Non-Disclosure Agreement (NDA).}

%The \netmob dataset represents a valuable new resource for mobility research, enabling studies on travel behavior, transport mode choice, activity patterns, teleworking, and more. It is made available to the community to foster methodological innovation and support research addressing societal and urban mobility challenges.
\end{abstract}

\section{Introduction}
%\jos{Add citations everywhere!}

The widespread adoption of mobile devices and location-aware technologies continues to generate large-scale, fine-grained data about human activities and movement patterns. Such data hold immense potential for the quantitative study of urban systems~\cite{urban1}, transportation planning~\cite{transportation1}, spatial inequality~\cite{spatial_ineq1, spatial_ineq2}, and the societal impacts of technological infrastructures~\cite{technological1}. In particular, mobility data have become instrumental in disciplines ranging from geography and sociology to computer science, public health, and economics, enabling the empirical analysis of phenomena such as commuting behavior~\cite{commuting1}, access to public services~\cite{public_service_access1}, environmental exposure~\cite{environmental_exp1}, or epidemic spread~\cite{epidemic1}.

Over the past decade, the research community has increasingly turned to mobile network data as a scalable alternative to traditional surveys, offering insights at temporal and spatial scales that were previously unattainable. Yet, despite its demonstrated value, access to high-quality mobility data remains the exception rather than the rule. Privacy concerns, proprietary restrictions, and the strategic value of data assets often prevent mobile network operators and digital platforms from sharing granular data, stifling innovation and hindering the reproducibility of scientific results.

To alleviate these limitations, open data challenges have played a pivotal role in fostering methodological progress and enabling large-scale empirical studies. Pioneering efforts such as the \textit{Data for Development} (D4D) challenges by Orange in collaboration with NetMob (2012--2014)~\cite{d4d_2013, d4d_2014}, the ITU Ebola challenge (2015)~\cite{itu_bigdata_epidemics}, the Telecom Italia Big Data Challenge (2014--2015)~\cite{Telecom_italia}, the Data for Refugees (D4R) challenge in Turkey (2018)~\cite{Data4R}, and the Future Cities Challenge supported by Foursquare at NetMob 2019~\cite{foursquare2018}, have collectively shaped a vibrant research landscape around anonymized mobility data.

More recently, the NetMob conference series has revived this tradition through curated data releases that set new standards of accessibility and scientific utility. \texttt{NetMob23}~\cite{netmob23} provided an extensive dataset of mobile application usage across 20 metropolitan areas in France, mapped at a 100$\times$100m spatial resolution and covering 68 popular mobile services. \texttt{NetMob24}~\cite{netmob2024} shifted the focus toward countries of the Global South, offering aggregated mobility statistics from Mexico, Colombia, Indonesia, and India, in line with Sustainable Development Goals and local policy needs.

For the 2025 edition, NetMob Data Challenge releases a unique \netmob dataset describing GPS-based travel records, described in this paper.
%In this paper, we introduce the \netmob dataset, released as part of the NetMob 2025 Data Challenge. 
This dataset offers a fundamentally different perspective on mobility, as it is derived not from passively collected network activity, but from a dedicated, high-resolution GPS-based mobility survey conducted in the Île-de-France region (Greater Paris area) between October 2022 and May 2023. It provides access to ground-truth data on the daily movements of \textbf{3,337 participants} over seven days, captured with a temporal resolution of \textbf{2--3 seconds}.

Each participant was equipped with a GPS tracking device %, \ali{(a BT-Q1000XT Bluetooth A-GPS eXtreme Travel Recorde)} 
and asked to maintain a travel diary~ -- either digitally or on paper~-- with all trajectories subsequently validated via phone interviews. The final dataset includes individual-level sociodemographic data, detailed trip annotations with modes and purposes, and raw GPS trajectories representing over 500 million location points. A calibrated statistical weighting mechanism is also provided to support inferences at the population scale.

The \netmob dataset offers a unique opportunity to study individual mobility in a large European metropolis with unprecedented granularity and behavioral accuracy. It supports a wide range of potential applications, including:
\begin{itemize}[leftmargin=*]
  \item The analysis of travel behavior across sociodemographic groups and household structures;
  \item The modeling of multimodal mobility and transport mode choices;
  \item The study of temporal rhythms of urban activity and teleworking patterns;
  \item The design and evaluation of trip inference algorithms and segmentation methods;
  \item The spatial analysis of flows, accessibility, and attendance at fine geographical scales;
  \item Investigations of mobility equity, sustainability, and service provision in dense urban environments.
\end{itemize}

By combining rich individual annotations with high-frequency location data, \netmob fills a critical gap in the mobility data landscape. It provides researchers with a robust, multi-layered dataset that preserves individual privacy while enabling in-depth analyses of urban mobility dynamics. The \netmob dataset will be made available to participants only upon acceptance of the challenge’s \textit{Terms and Conditions} and the signing of a Non-Disclosure Agreement (NDA)

To promote rapid dataset exploitation, this paper presents the data collection methodology, the structure of the released databases, the anonymization and cleaning procedures, and the empirical mobility characterization of the data, aiming to support its future use across disciplines and domains. The dataset characterization focuses on individual-level analytics and complements the aggregated and socioeconomic analyses presented in EMG2~\cite{EMG-slides} and discussed in~\cite{EMG-introduction}.

\section{Data Collection}
Understanding everyday mobility in Île-de-France is essential for informed policy design in transportation and urban planning. However, acquiring accurate and up-to-date data is a persistent challenge, typically requiring large-scale and costly surveys. To address this, \textit{L’Institut Paris Region}, supported by a consortium of public and private stakeholders, launched the EMG 2023 (Enquête Mobilité par GPS) initiative. This effort aimed to capture a recent, high-resolution snapshot of regional mobility through passive data collection, complementing traditional methods like the "Enquête Globale Transport" (EGT).

\vspace{1em}
\noindent\textbf{Survey objectives and scope.}
The EMG 2023 survey took place between October 2022 and May 2023 and focused on residents aged 16 to 80 in Île-de-France. Aiming to provide a snapshot of the mobility of Île-de-France residents, the survey excluded non-residents, tourists, and individuals who are immobile due to health reasons. A total of \textbf{3,337 participants} were recruited using a multichannel strategy combining quota sampling and randomized draws to ensure a diverse and representative sample. The survey goal was to observe mobility behavior over a full week to capture multimodality, weekday–weekend differences, and daily travel patterns to diversify the understanding of mobility among Île-de-France residents a year and a half after the end of the COVID-19 pandemic.

\vspace{1em}
\noindent\textbf{Data collection protocol.}
Each participant was equipped with a dedicated GPS tracking device~-- BT-Q1000XT Bluetooth{\tiny®} A-GPS eXtreme Travel Recorder~-- recording locations every 2 to 3 seconds during movement. Crucially, the device only logged positions when motion was detected, meaning that \textit{gaps in the data correspond to stationary periods or absence of mobility, rather than signal loss \cite{qstarz_manual}}. Participants were instructed to carry the device throughout the day for \textbf{seven consecutive days}. 

In parallel, participants filled out a \textbf{daily travel diary} (either digitally or on paper), reporting trip purposes, used transport modes, and context. Upon completion of the week, each participant was contacted for a \textbf{follow-up phone interview} in which an interviewer verified or corrected the trip information inferred from the GPS traces using the logbook data.

\vspace{1em}
\noindent\textbf{Trip detection and data processing.}
GPS traces were pre-processed by the mobility analytics provider \textit{Hove}, who applied proprietary algorithms to identify displacements, segment trips, and label them with preliminary metadata (e.g., duration, mode). The manual verification phase ensured the reliability of detected trip boundaries and semantic annotations. This hybrid protocol~-- combining passive sensing, self-reporting, and human validation~-- makes EMG 2023 notably robust and behaviorally rich.

\vspace{1em}
\noindent\textbf{Temporal and contextual coverage.}
The data collection spanned \textbf{20 weeks}, from mid-October 2022 to mid-May 2023, and includes a variety of calendar contexts such as weekdays, weekends, public holidays, school holidays, and strike days. These special days were not excluded but are explicitly annotated in the dataset, allowing users to identify and filter atypical mobility conditions. This enables nuanced analyses of temporal rhythms and behavioral variability under different societal or institutional circumstances.

% \vspace{1em}
% \noindent\textbf{Comparison with traditional surveys.}
% Table~\ref{tab:methodological-differences} summarizes the key methodological differences between EMG 2023 and the EGT 2018 survey. EMG introduces several improvements, including automatic round-trip capture, high spatial accuracy, a full week of observations, and hybrid annotation strategies.

% \begin{table}[h!]
% \centering
% \small
% \begin{tabularx}{\textwidth}{|X|X|}
% \hline
% \textbf{EGT (2018)} & \textbf{EMG (2023)} \\
% \hline
% Face-to-face/phone interviews & GPS tracking + logbook + phone verification \\
% \hline
% 1 day (randomly assigned) & 7 consecutive days \\
% \hline
% Random sampling & Quota + random draw recruitment \\
% \hline
% Ages 5+ & Ages 16–80 (resident population) \\
% \hline
% Recall-based reporting & Passive high-frequency logging \\
% \hline
% Limited round-trip reporting & Full round-trip capture \\
% \hline
% Spatial resolution: map scale & Resolution down to postal address \\
% \hline
% Trip routes not collected & Raw GPS traces available \\
% \hline
% 7k–9k individuals/year & 3,337 individuals (1/2,700 ratio) \\
% \hline
% Survey over 10 months & 5-month window \\
% \hline
% Detailed questionnaire & Simplified diary + annotations \\
% \hline
% \end{tabularx}
% \caption{Comparison of methodological approaches between EGT 2018 and EMG 2023.}
% \label{tab:methodological-differences}
% \end{table}

\section{Dataset Overview and Anonymization}
%\section{Dataset Overview and Anonymization}

\begin{figure}[h!]
    \centering
    \includegraphics[width=\linewidth]{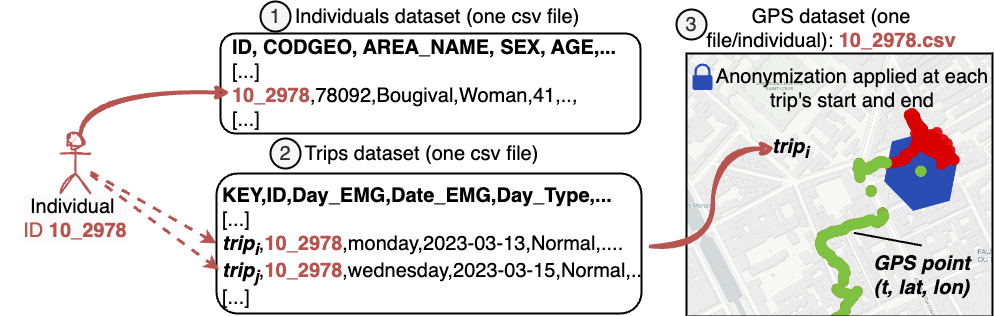}
    \caption{Structure of the EMG dataset for a single participant (\texttt{10\_2978}). Each participant is represented in the Individuals dataset (1 row), the Trips dataset (multiple rows over the week), and a personal GPS file. GPS traces are anonymized by removing points outside of recorded trips (red dots), and blurring only the start and end of each trip (green dots) using the centroid of the corresponding H3 cell (resolution 10). The in-trip GPS points are preserved at full temporal and spatial granularity.}
    \label{fig:emg-structure}
\end{figure}

The EMG dataset released in the context of the \netmob{} challenge consists of three primary components, each representing different layers of information about the mobility of 3,337 residents in the Île-de-France region. These datasets are individually rich but designed to be interoperable through a shared participant identifier. Fig.~\ref{fig:emg-structure} illustrates the structure of the dataset and the anonymization process applied to the GPS records.

\subsection*{Database structure}

The dataset is composed of three distinct but linkable CSV-based databases:

\begin{itemize}[leftmargin=*]
  \item \textbf{Individuals database} (3,337 rows): Contains sociodemographic and household-level attributes for each participant (e.g., age, sex, residence code, education, employment, car ownership).
  
 \item \textbf{Trips database} ($n \approx 80{,}697$): Contains one row per validated trip made by a participant. Each row includes the date, type of day (e.g., normal, strike, holiday), origin and destination information, transport modes (including multimodal combinations), duration, and purpose of the trip. Days without any trips are also recorded with a corresponding row indicating the absence of mobility.

  \item \textbf{GPS traces database} ($\approx$500 million points): Contains raw high-frequency GPS points recorded every 2–3 seconds for each participant over seven consecutive days. Each participant has a dedicated file named after their anonymized ID.
\end{itemize}

All databases are keyed by a unique \texttt{ID} field, enabling cross-referencing. For example, individual-level attributes (e.g., sex or household structure) can be joined to the trip records to allow mobility analysis by population group. Conversely, trip-level indicators such as total travel time can be aggregated and reintegrated into the Individuals database to build behavioral profiles.

\subsection*{Anonymization and cleaning procedures}

To ensure privacy and comply with GDPR regulations, an extensive anonymization pipeline was applied to the GPS traces. The process was carefully designed to preserve the analytical utility of the data while removing sensitive or non-informative elements that could compromise user confidentiality. The key steps of this pipeline are summarized below:

\begin{itemize}[leftmargin=*]
    \item \textbf{ID pseudonymization:} A unique, consistent identifier was assigned to each participant and used across the three datasets. No personal identifiers (such as names or addresses) are present in the released data.

    \item \textbf{General cleaning:} Devices with incomplete or anomalous records were excluded from the GPS traces. Trip annotations were harmonized, and missing values were manually reviewed during the post-survey validation phase. Notably, while the Individuals and Trips databases reference 3,337 participants, only 3,320 of them have corresponding GPS records. The 17 users with missing or incomplete GPS data were retained, as their sociodemographic profiles and weights remain useful for independent analyses. These individuals are flagged with a \texttt{GPS\_RECORD} boolean column in the Individuals file.

    \item \textbf{Removal of non-trip data:} All GPS points not belonging to validated trips—typically corresponding to stationary phases or short, non-representative intra-location movements—were discarded. These removed points are visually illustrated as {red dots} in Fig. ~\ref{fig:emg-structure}.

    \item \textbf{Spatial blurring:} To protect the privacy of frequently visited places (e.g., home, workplace), only the start and end of each trip are spatially blurred. The first and last GPS points of each trip, along with any subsequent points falling within the same spatial cell, are mapped to the centroid of the corresponding H3 hexagonal grid cell at resolution 10, using the \href{https://h3geo.org}{\texttt{h3geo}} system. This process hides approximately the first and last 50 to 100 meters of each trip. Importantly, this anonymization is applied independently to each trip—there is no spatial consistency in the cell assignment across different days or trips. The intermediate GPS points, shown as unblurred paths in Fig. ~\ref{fig:emg-structure} (green), are preserved at full spatiotemporal resolution to retain the integrity of the trajectory for research use.
\end{itemize}

This anonymization strategy ensures that locations such as homes, workplaces, or personal points of interest are protected, while the internal structure of trips remains usable for trajectory, transport, and behavioral analyses.

\section{Database Structure and Variables}
%\section{Variable Dictionaries}
This section describes the structure and content of the three core datasets provided in the EMG study. For each dataset, we present the main variables and their semantics. In addition, we summarize the methodology used to compute statistical weights, allowing extrapolation from the sample to the population of Île-de-France. A more detailed description of the datasets structure and weighting procedure is available in the official dataset documentation \cite{netmob25slides}.

\subsection{Individual Database}
The \textbf{Individuals dataset} contains one row per participant, including demographic, socioeconomic, and mobility-related information.

%\alic{considering the anonymization of last section (...ensures that locations such as homes, workplaces, or personal points of interest are protected), what the \texttt{TYPE\_HOUSE} shows for anonymized places? it would be worth mentioning something on this here. }

\begin{table}[h!]
\centering
\scriptsize
\begin{tabularx}{\textwidth}{lX}
\texttt{ID} & Unique individual identifier \\
\texttt{CODGEO} & INSEE code of municipality of residence \\
\texttt{AREA\_NAME} & Municipality name \\
\texttt{SEX} & Sex of the individual (Woman/Man) \\
\texttt{AGE} & Age in years \\
\texttt{DIPLOMA} & Highest diploma obtained, from no diploma to doctoral level \\
\texttt{PRO\_CAT} & Socio-professional category (1--8) \\
\texttt{TYPE\_HOUSE} & Type of household or cohabitation status (e.g., In a shared appartment)\\
\texttt{NBPERS\_HOUSE} & Number of persons in household \\
\texttt{NB\_10}, \texttt{NB\_11\_17}, \texttt{NB\_18\_24}, \texttt{NB\_25\_64}, \texttt{NB\_65} & Number of household members in each age range \\
\texttt{PMR} & Mobility impairment indicator \\
\texttt{DRIVING\_LICENCE} & Possession of a driving licence for cars (B) \\
\texttt{NB\_CAR} & Number of cars in household \\
\texttt{TWO\_WHEELER}, \texttt{BIKE}, \texttt{ELECT\_SCOOTER} & Number of other mobility devices \\
\texttt{NAVIGO\_SUB}, \texttt{IMAGINER\_SUB}, \texttt{OTHER\_SUB\_PT} & Public transport subscriptions \\
\texttt{BIKE\_SUB}, \texttt{NSM\_SUB} & Other mobility service subscriptions \\
\texttt{WEIGHT\_INDIV} & Statistical weighting coefficient \\
\texttt{GPS\_RECORD} & Boolean indicating whether the individual has associated GPS records \\
\end{tabularx}
\caption{Key variables in the Individuals dataset.}
\end{table}

\subsection{Trip Database}
Each row in the \textbf{Trips dataset} corresponds to a recorded trip or day without mobility.

\begin{table}[h!]
\centering
\scriptsize
\begin{tabularx}{\textwidth}{lX}
\texttt{KEY} & Unique trip identifier \\
\texttt{ID} & Individual identifier \\
\texttt{Day\_EMG}, \texttt{Date\_EMG} & Day and date of record \\
\texttt{Day\_Type} & Contextual day type: normal, holiday, strike, etc. \\
\texttt{ID\_Trip\_Days} & Trip number of the day \\
\texttt{No\_Traces}, \texttt{No\_Trip} & Flags for GPS recording or travel occurrence \\
\texttt{Outside\_IDF}, \texttt{Type\_Trip\_OD} & Regional indicators for trip origin/destination \\
\texttt{Area\_O}, \texttt{Area\_D} & Origin and destination municipalities \\
\texttt{Code\_INSEE\_O}, \texttt{Code\_INSEE\_D} & Corresponding INSEE codes \\
\texttt{Zone\_O}, \texttt{Zone\_D} & Urban zones: Paris, inner/outer suburbs \\
\texttt{Date\_O}, \texttt{Time\_O}, \texttt{Date\_D}, \texttt{Time\_D} & Timestamps of trip start and end \\
\texttt{Duration} & Duration in minutes \\
\texttt{Purpose\_O}, \texttt{Purpose\_D} & Trip purposes at origin and destination \\
\texttt{Main\_Mode} & Main mode used (e.g., car, metro, bike) \\
\texttt{Mode\_1} to \texttt{Mode\_5} & All modes used for multimodal trips \\
\texttt{Weight\_Day} & Daily statistical weight \\
\end{tabularx}
\caption{Key variables in the Trips dataset.}
\end{table}

\subsection{GPS dataset}

The \textbf{GPS dataset} consists of individual CSV files containing raw location traces collected from participants. Each file corresponds to one participant and is named using their unique \texttt{ID}. These traces were recorded using a professional-grade GNSS device and are time-aligned with the validated trips listed in the trip database, enabling detailed trajectory reconstruction.

Each row in a GPS file represents a single location sample, including the following fields:

\begin{itemize}[leftmargin=*]
    \item \texttt{UTC TIMESTAMP}: Coordinated Universal Time (UTC) indicating when the sample was collected.
    \item \texttt{LOCAL TIMESTAMP}: Local timestamp with time zone offset applied.
    \item \texttt{LATITUDE}, \texttt{LONGITUDE}: Geographic coordinates in signed decimal degrees.
    \item \texttt{VALID}: GPS fix quality, taking one of four values — \texttt{SPS}, \texttt{DGPS}, \texttt{Estimated}, or \texttt{NO FIX}. This field helps identify potential signal degradation due to environmental factors such as urban canyons, tunnels, or device initialization phases. For example, an \texttt{Estimated} value typically indicates degraded GPS quality, signaling possible uncertainty in the position.
    \item \texttt{SPEED}: Ground speed calculated using Doppler shift rather than positional differencing. While \texttt{DGPS} significantly improves position accuracy by correcting satellite signal errors (e.g., atmospheric delays), velocity is independently estimated using the Doppler effect. This method analyzes frequency shifts in the GNSS signals caused by relative motion, yielding velocity accuracy on the order of a few centimeters per second. In summary the \textbf{position accuracy} is enhanced by DGPS corrections while the \textbf{speed accuracy} depends on Doppler-based estimations and the quality of satellite signal reception.
\end{itemize}

\subsection{Weight construction}

To ensure representativeness, calibration weights were computed independently for individuals and trips using pairwise marginal distributions derived from census statistics. These weights allow extrapolation from the observed sample to the Île-de-France population.

\paragraph{Individual weights.}  
% Each participant is assigned a weight (\texttt{WEIGHT\_INDIV}) so that aggregated statistics match population distributions across the following dimensions:
% \begin{itemize}[leftmargin=*]
%     \item Department × Age group (16–25, 26–45, 46–65, 66–80);
%     \item Department × Sex;
%     \item Department × Socio-professional category;
%     \item Department × Number of cars;
%     \item Department × Household size;
%     \item Department × Diploma level.
% \end{itemize}
Each participant is assigned a weight (\texttt{WEIGHT\_INDIV}) representing how many individuals in the Île-de-France region share the same sociodemographic profile. This profile is defined by the cross-tabulation of several variables: department of residence (8 departments), age group (16–25, 26–45, 46–65, 66–80), sex (male, female), socio-professional category (craftsmen, executives, intermediate professions, employees and workers, retirees and other inactives), number of cars in the household (0, 1, or 2+), household size (1, 2, 3, or 4+ people), and highest diploma obtained (lower or upper secondary, Bac+2, Bac+5 or doctorate).

The weights are calibrated such that their sum matches the total population. For instance, the number of executives can be estimated by summing the weights of all individuals with \texttt{PRO\_CAT = 2}. Similarly, to estimate the number of individuals across all age groups residing in a specific department, one can sum the weights of all users meeting that criterion. 

%\alic{Josiane, i think your example is better. we could add it here.}

\paragraph{Trip weights.}  
%Each reported day in the trips database is weighted via the variable \texttt{Weight\_Day}. 
Each reported day in the trips database is weighted using the variable \texttt{Weight\_Day}, which reflects how many individuals from the regional population share the same sociodemographic profile and trip-day characteristics. The weighting combines the same set of sociodemographic variables used for individuals—department, age group, sex, socio-professional category, household size, number of cars, and diploma level—along with the day of the week (Monday to Sunday).
This calibration accounts for the same sociodemographic criteria as above, plus the day of the week, to reflect variations in daily travel patterns. Usage examples include:
\begin{itemize}[leftmargin=*]
    \item Weighting all trips across the week directly;
    \item Weighting weekday trips only (multiply by $7/5$);
    \item Weighting a specific day (e.g., Monday) by multiplying by 7.
\end{itemize}

\section{Exploratory Dataset Characterization}
%\section{Exploratory Dataset Characterization}

This section provides an empirical overview of the EMG dataset by exploring individual-level characteristics, trip statistics, transport mode usage, and spatial flow patterns. Our goal is to assess both the coverage and representativity of the sample, as well as to highlight key behavioral trends within the population. We rely on weighted distributions to ensure that our statistics align with the known demographics of the Île-de-France region, following the calibration procedure detailed in the dataset documentation \cite{netmob25slides}.

\subsection{Individual-Level Statistics}

\begin{figure}[h]
    \centering
    \begin{minipage}[b]{0.48\linewidth}
        \centering
        \includegraphics[width=0.8\linewidth]{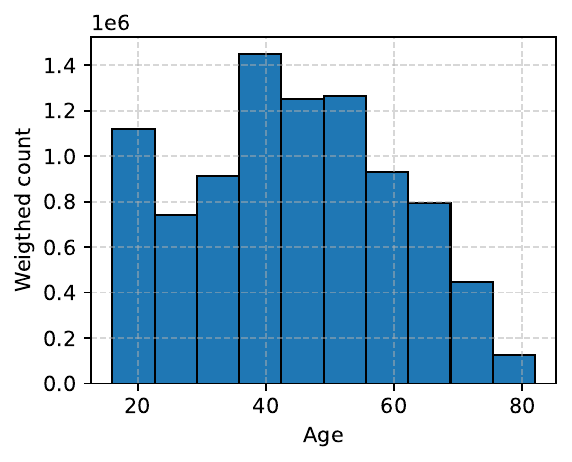}
        \caption{Distribution of participants' age.}
        \label{fig:age-distribution}
    \end{minipage}
    \hfill
    \begin{minipage}[b]{0.48\linewidth}
        \centering
        \includegraphics[width=0.8\linewidth]{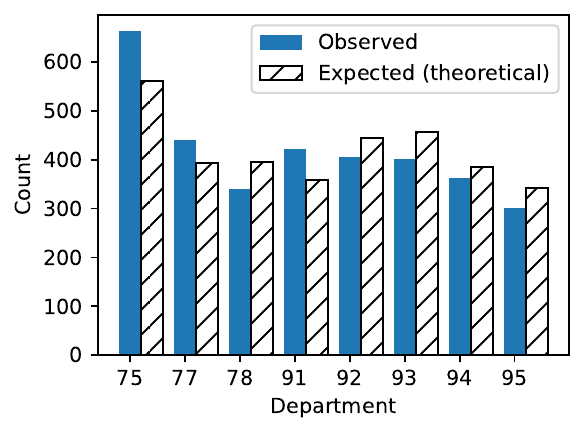}
        \caption{Observed vs. expected department-level distribution.}
        \label{fig:dept-observed-expected}
    \end{minipage}
\end{figure}

\paragraph{Age and sex distribution.}  
The survey targeted participants aged between 16 and 81 years old. Fig. ~\ref{fig:age-distribution} shows the weighted distribution of age in the sample, highlighting good alignment with the reference population in Île-de-France (based on INSEE 2021 data).

In terms of sex, the dataset includes 1,836 women and 1,501 men, yielding a sex ratio slightly skewed toward female participants (55\%). After applying the individual weights, the sample represents approximately 4,708,201 women and 4,335,886 men in the Île-de-France population, bringing the sex ratio closer to parity and aligning the dataset with regional demographic statistics.

\paragraph{Department of residence.}  
We also examine the geographical distribution of participants by department. Fig. ~\ref{fig:dept-observed-expected} contrasts observed and expected counts based on INSEE data. Paris (75), Essonne (91), and Seine-Saint-Denis (93) appear overrepresented, while others are underrepresented.

\begin{figure}[h]
    \centering
    \begin{subfigure}{0.48\textwidth}
        \centering
        \includegraphics[width=0.7\linewidth]{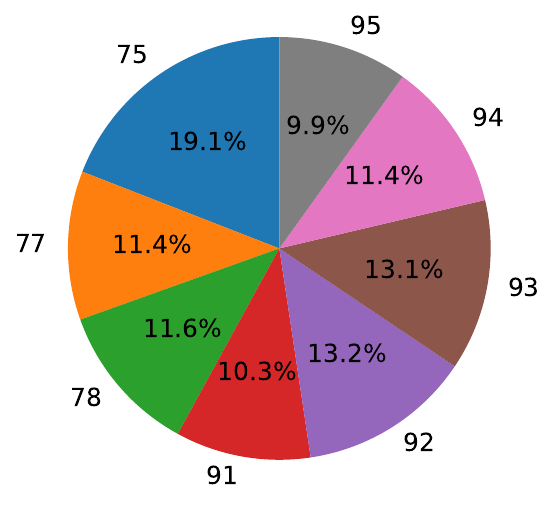}
        \caption{Pie chart of weighted department distribution}
        \label{fig:dept-pie}
    \end{subfigure}
    \hfill
    \begin{subfigure}{0.5\textwidth}
        \centering
        \includegraphics[width=0.7\linewidth]{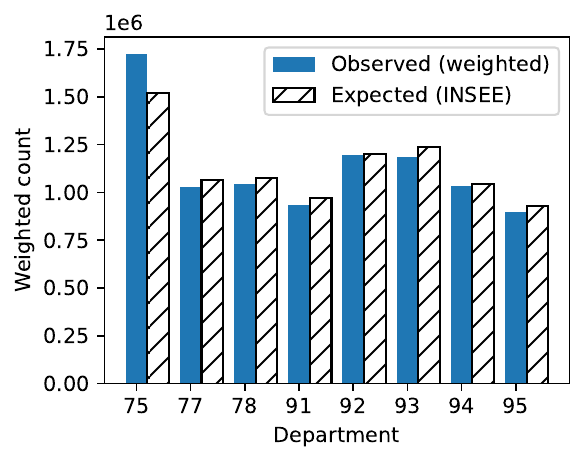}
        \caption{Bar chart of weighted department distribution}
        \label{fig:dept-bar}
    \end{subfigure}
    \caption{Weighted distribution of the population by department in the dataset}
    \label{fig:dept-weighted}
\end{figure}

To statistically assess the difference, a chi-squared goodness-of-fit test yields a statistic of 60.45 (7 degrees of freedom, $p = 1.2 \times 10^{-9}$), confirming a significant deviation from proportional representation. As a corrective, post-stratification weights are applied to each individual to restore representativity (Fig. ~\ref{fig:dept-weighted}).

% \begin{figure}[h]
%     \centering
%     \includegraphics[width=6cm]{Figures/piechart.pdf}
%     \includegraphics[width=6cm]{Figures/barchartWEIGHT.pdf}
%     \caption{Weighted distribution of population by department.}
%     \label{fig:dept-weighted}
% \end{figure}

\subsection{Trip-Level Statistics}

This section analyzes the volume, temporal structure, and spatial dynamics of trips recorded during the collection period, including inter-departmental and inter-zonal flows, trip durations, and their variation across transport modes.

\paragraph{Temporal coverage and trip counts.}

\begin{figure}[!htb]
    \centering
    \begin{tiny}
    \subfloat[Per date]{\includegraphics[width=.49\linewidth]{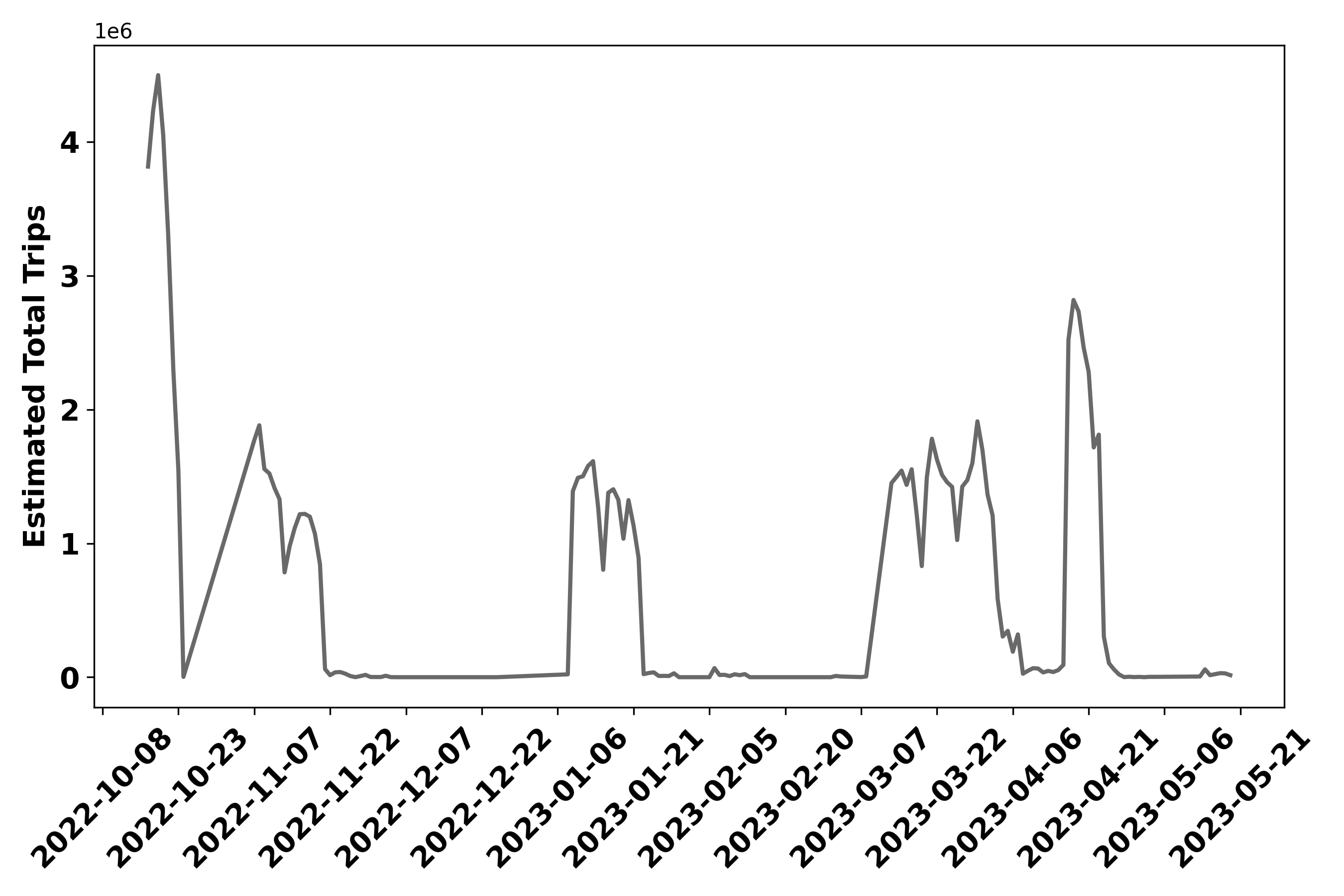}}\hspace*{\fill}
    \subfloat[Per hour]{\includegraphics[width=.49\linewidth]{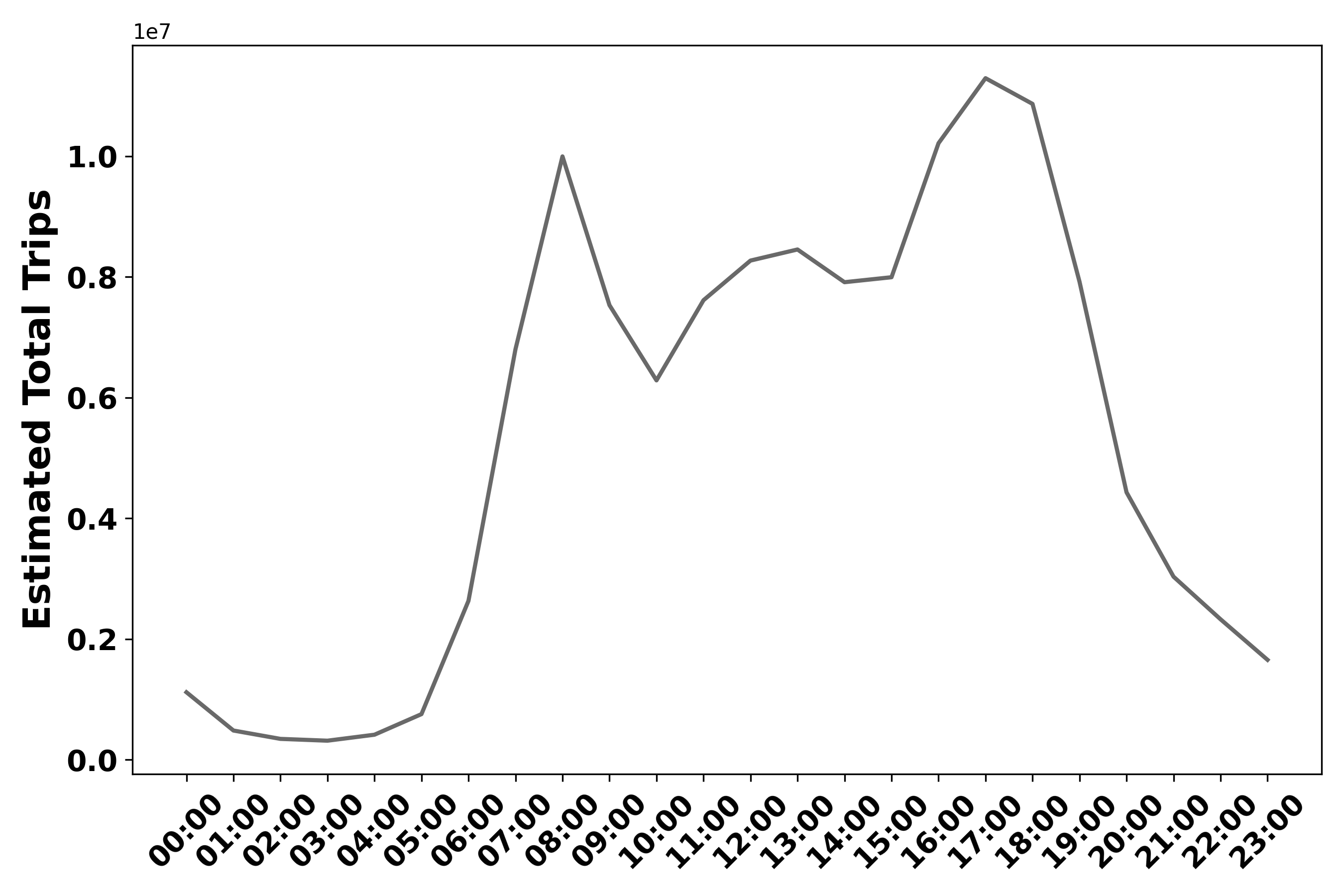} }
    \end{tiny}
    \caption{Estimated weighted number of trips per date and per hour.}
    \label{fig:trip_count}
\end{figure}

The dataset covers the period from October 17, 2022, to January 15, 2023. However, participants did not record trips continuously throughout this interval.
%Fig. ~\ref{fig:numberofrides} shows the evolution of the total number of detected trips per day throughout the observation period.
Fig.~\ref{fig:trip_count} presents the weighted number of trips aggregated by day and by hour. These are computed by summing the values of the \texttt{Weight\_Day} column from the \texttt{trips\_dataset}, aggregating over unique trips for each time interval.
Fig.~\ref{fig:trip_count}(a) shows the estimated daily number of trips throughout the observation period. Notable spikes are observed at the start of the data collection, between 2022-12-22 and 2023-01-10, and again from 2023-03-25 to 2023-04-15.
Fig.~\ref{fig:trip_count}(b) illustrates the distribution of trip start times by hour of the day. As expected, trip activity is minimal during nighttime and begins increasing around 05:00. Two distinct peaks occur at 07:00 and 17:00, corresponding to typical commuting periods for work, school, or other daily routines.

%To compute the number of trips shown in Fig.~\ref{fig:trip_count}, we consider the positions $P = \{p_1, p_2, ..., p_n\}$ of each individual from the \texttt{gps\_dataset} directory. The number of trips for each IRIS $s$ is calculated by mapping the routes that pass through $s$, identifying the trip weights from the \texttt{Weight\_Day} column in the \texttt{trips\_dataset}, and summing the weights of all unique trips.

% \begin{figure}[!h]
% \begin{center}
%     \includegraphics[width=0.6\linewidth]{Figures/numberofrides.pdf}
% \end{center}
% \caption{Number of trips observed per day (unweighted and weighted)}
% \label{fig:numberofrides}
% \end{figure}

\begin{figure}[H]
    \centering
    \begin{tiny}
    \subfloat[Trips per user per date.]{\includegraphics[width=.49\linewidth]{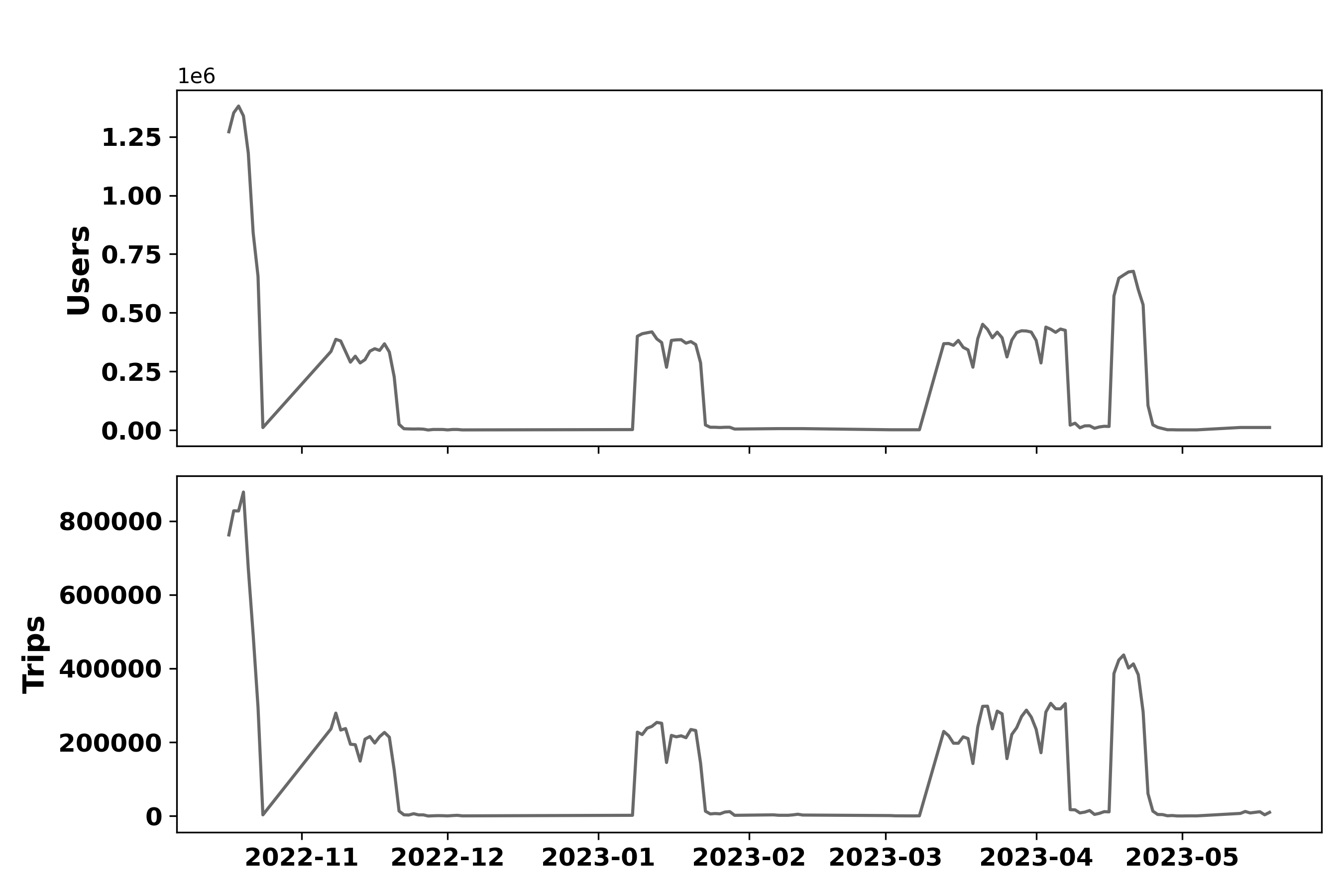}}\hspace*{\fill}
    \subfloat[Trips per user per weekday.]{\includegraphics[width=.49\linewidth]{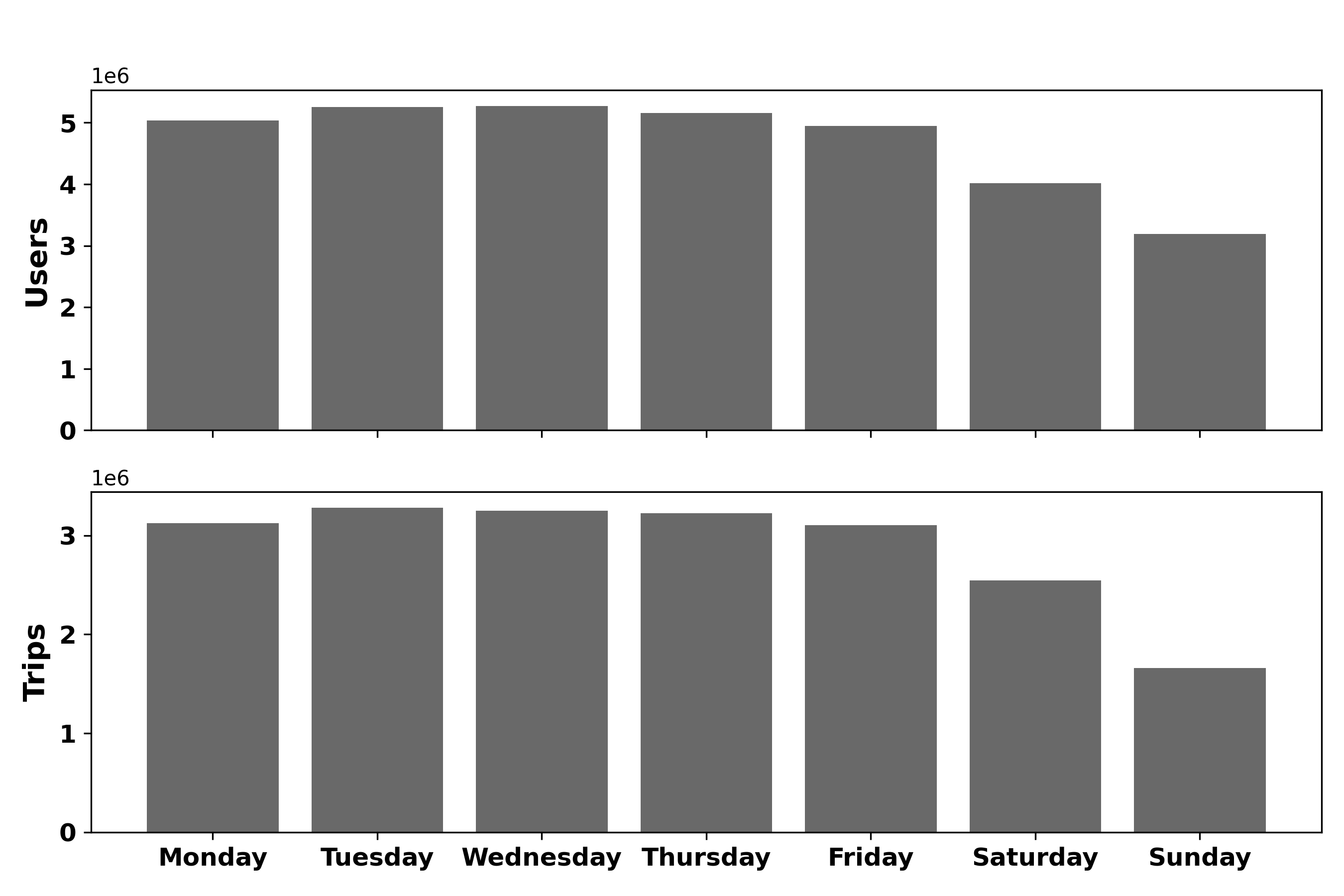}}
    \end{tiny}
    \caption{Number of users and trips per date and per day of the week.}
    \label{fig:total_trips_user}
\end{figure}

A more detailed view of the weighted number of individuals and trips by date and weekday is provided in Fig.~\ref{fig:total_trips_user}. Fig.~\ref{fig:total_trips_user}(a) shows the estimated number of individuals and their corresponding trips per date. These estimates are derived from the \texttt{trips\_dataset}, after excluding entries where individuals did not travel (\texttt{ID\_Trip\_Days} = \texttt{No\_Trip}), lacked recorded traces (\texttt{ID\_Trip\_Days} = \texttt{No\_Traces}), or were not associated with any displacement in the \texttt{gps\_dataset}. The weighted number of trips per individual per date is calculated by multiplying the raw trip count by the corresponding day weight: \texttt{Weighted\_Trip\_Count = nTrips * Weight\_Day}. The number of individuals per day is then estimated by aggregating their respective weights (\texttt{WEIGHT\_INDIV}) by date.

Fig.~\ref{fig:total_trips_user}(b) presents the same metrics aggregated by weekday. Trip frequencies display a strong weekly pattern, with both the number of individuals and the trip volume declining sharply on Saturdays and Sundays—particularly on Sundays, which consistently exhibit the lowest mobility. This reduction suggests decreased commuting activity and routine travel during weekends, in line with expected behavioral patterns.

%Fig.~\ref{fig:total_trips_user}(b) illustrates these metrics grouped by day of the week. A marked drop in both individual counts and trip volume is observed on Saturdays and Sundays, highlighting reduced mobility likely due to less work- or school-related travel during weekends.

\begin{figure}[!htb]
    \centering
    \includegraphics[width=0.5\textwidth]{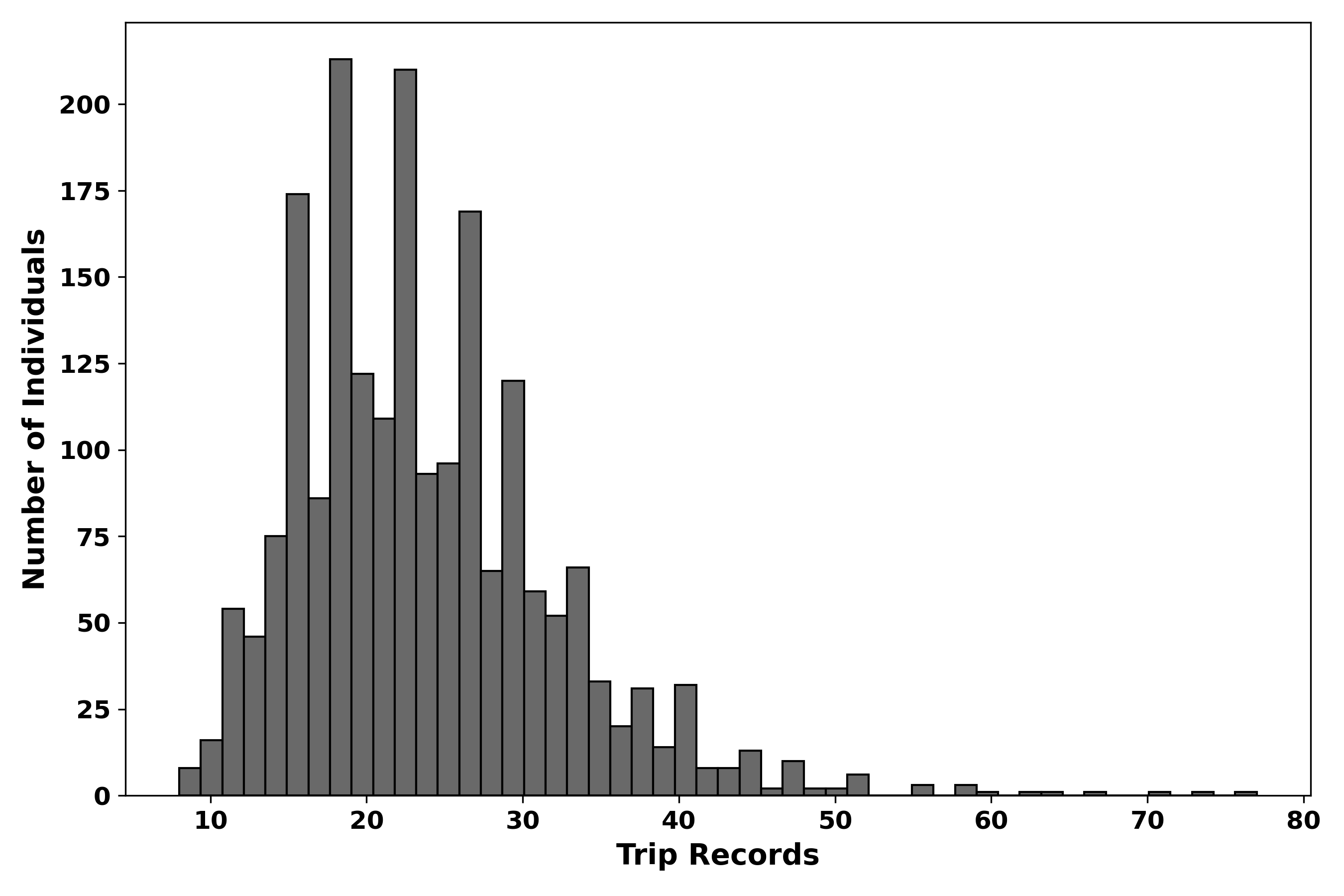}
    \caption{Distribution of trip records per individual.}
    \label{fig:trip_records_user}
\end{figure}

% \begin{figure}[!h]
% \centering
% \begin{subfigure}{0.45\textwidth}
%     \includegraphics[width=\linewidth]{Figures/rideweek.pdf}
%     \caption{Unweighted}
% \end{subfigure}
% \hfill
% \begin{subfigure}{0.45\textwidth}
%     \includegraphics[width=\linewidth]{Figures/rideweekweighted.pdf}
%     \caption{Weighted}
% \end{subfigure}
% \caption{Average number of trips by day of the week}
% \label{fig:rideweek}
% \end{figure}
\paragraph{Distribution of trip records per individual.}
Fig.~\ref{fig:trip_records_user} displays the distribution of the number of trip records per individual. Only individuals with corresponding GPS files in the \texttt{gps\_dataset} are considered, ensuring the availability of valid mobility traces. Each record corresponds to an entry in the \texttt{trips\_dataset}, including those labeled as \texttt{No\_Trip} or \texttt{No\_Traces}. The distribution reveals that the majority of individuals have approximately 23 records.

%Each record corresponds to a date in the \texttt{Date\_EMG} column of the \texttt{trips\_dataset}, with one entry per day when a trip was detected. The number of records per individual is determined by counting the unique dates associated with each individual, identified by the \texttt{ID} column. Individuals without a corresponding displacement file in the \texttt{gps\_dataset} are excluded, as they lack usable mobility traces. This count includes all valid entries, even those labeled \texttt{No\_Trip} or \texttt{No\_Traces}, provided a corresponding file is present in the displacement dataset.

% \paragraph{Number of trips per individual.}
% Fig.~\ref{fig:trip_records_user} illustrates the distribution of the number of records per individual. Most individuals have, on average, 23 records. To obtain this count, each entry in the \texttt{Date\_EMG} column (where each date represents one record) from the \texttt{trips\_dataset} file is considered. The number of records is calculated per individual, identified by unique values in the \texttt{ID} column. Individuals who do not have a corresponding file in the \texttt{gps\_dataset} folder are excluded from the count, as they have no displacement data available. \jos{Number of records inside the displacement files (including the No\_trips or No\_traces)}

\begin{figure}[H]
\centering
\begin{subfigure}{0.45\textwidth}
    \includegraphics[width=0.9\linewidth]{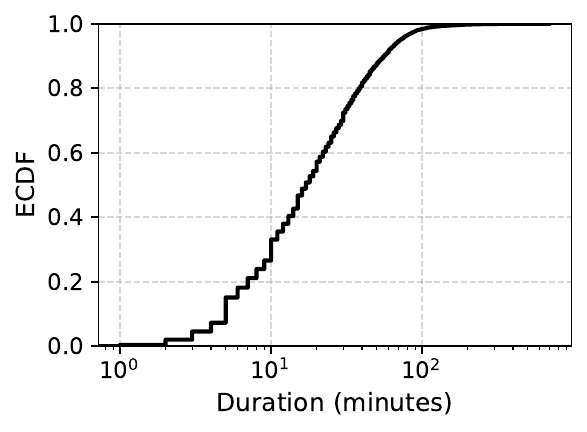}
    \caption{Empirical CDF}
\end{subfigure}
\hfill
\begin{subfigure}{0.5\textwidth}
    \includegraphics[width=0.8\linewidth]{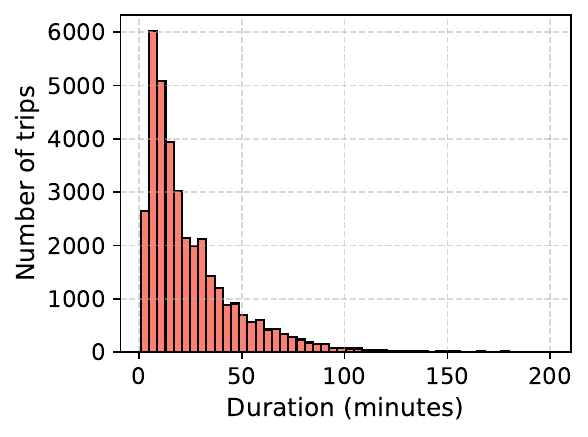}
    \caption{Histogram of trip durations  (capped at 200 min)}
\end{subfigure}
\caption{Distribution of trip durations}
\label{fig:tripdurations}
\end{figure}

\paragraph{Trip duration analysis.}
The distribution of trip durations is analyzed using both a histogram and an empirical cumulative distribution function (ECDF). As shown in Fig.~\ref{fig:tripdurations}, the average trip duration is approximately 30 minutes (mean = 29.97, SD $\approx$ 28 minutes), reflecting substantial variability across trips.

\begin{figure}[H]
\begin{center}
    \includegraphics[width=\linewidth]{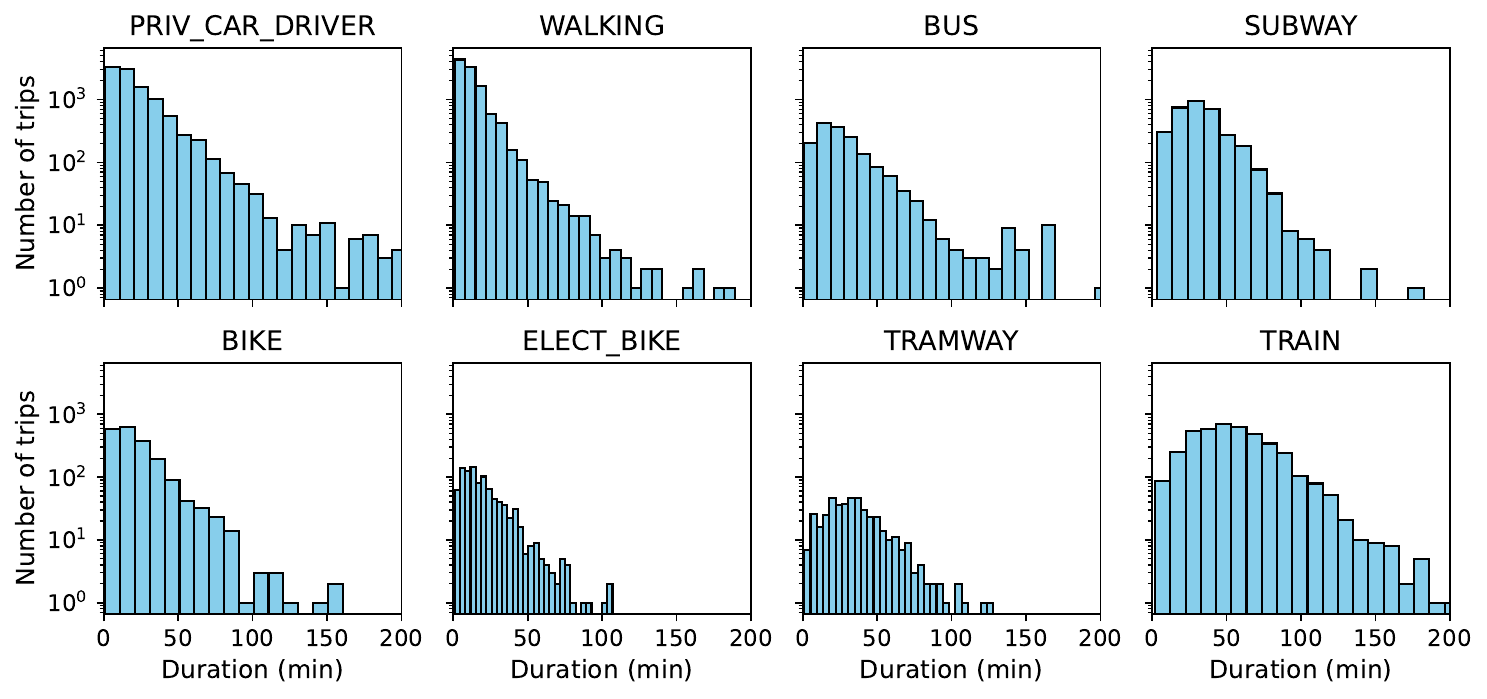}
\end{center}
\caption{Distribution of trip durations by transportation mode (capped at 200 min)}
\label{fig:durbytransport}
\end{figure}

Trip durations differ markedly depending on the primary transportation mode, as indicated by the \texttt{Main\_Mode} column. Fig.~\ref{fig:durbytransport} displays the distribution of trip durations for the eight most frequently used modes, highlighting mode-specific temporal characteristics.


\paragraph{Mobility between administrative zones.}

A key mobility variable concerns the flows between three major zones of Île-de-France:
\begin{itemize}[leftmargin=*]
  \item {Paris (intra-muros), i.e., department 75};
  \item {Inner suburbs (IS)} — departments 92, 93, 94;
  \item {Outer suburbs (OS)} — departments 77, 78, 91, 95.
\end{itemize}

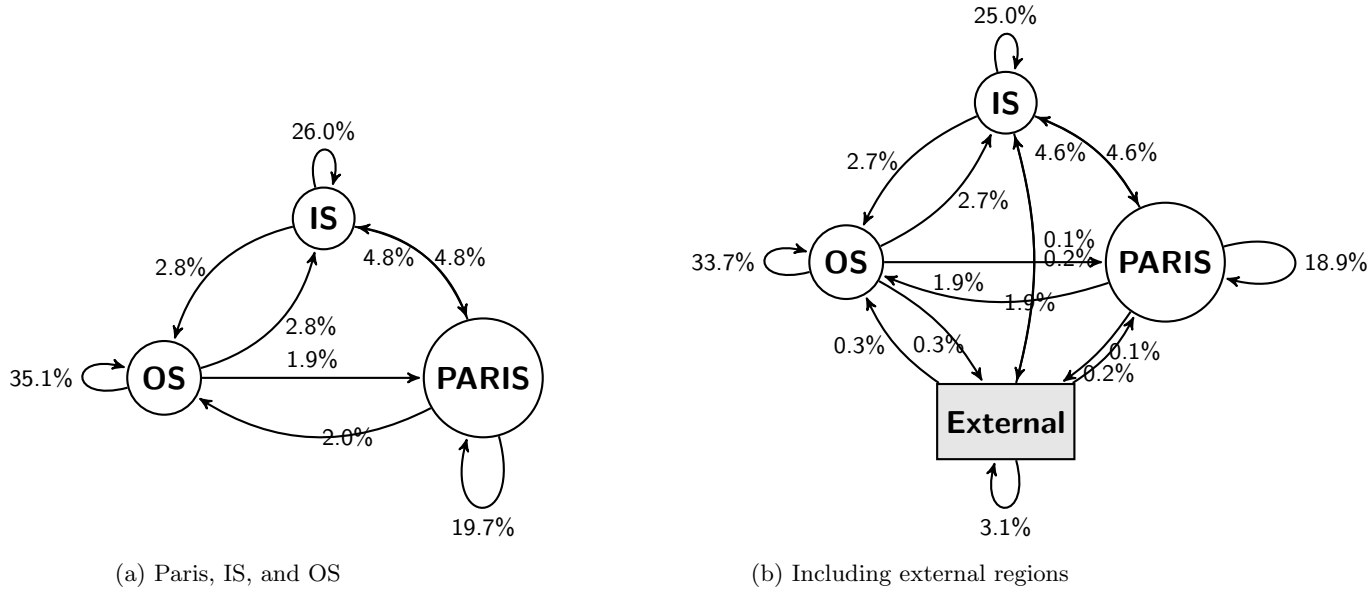
\begin{figure}[H]
\centering
\begin{subfigure}{0.4\textwidth}
\centering
\begin{tikzpicture}[->,>=stealth',shorten >=1pt,auto,node distance=3cm,
    thick,main node/.style={circle,draw,font=\sffamily\large\bfseries}]
    
    \node[main node] (IS)    {IS};
    \node[main node] (OS) [below left of=IS]    {OS};
    \node[main node] (PARIS) [below right of=IS] {PARIS};
    
    \path[every node/.style={font=\sffamily\small}]
    (IS) edge [loop above]             node {26.0\%} (IS)
         edge [bend right]             node[left]  {2.8\%}  (OS)
         edge [bend left]              node[right] {4.8\%}  (PARIS)
    (OS) edge [bend right]             node[right] {2.8\%}  (IS)
         edge [loop left]              node {35.1\%} (OS)
         edge                          node {1.9\%}  (PARIS)
    (PARIS) edge [bend right]          node[left]  {4.8\%}  (IS)
           edge [bend left]           node[right] {2.0\%}  (OS)
           edge [loop below]          node {19.7\%} (PARIS);
\end{tikzpicture}
\caption{Paris, IS, and OS}
\label{fig:interzonalflows}
\end{subfigure}
\hfill
\begin{subfigure}{0.4\textwidth}
\centering
\begin{tikzpicture}[->,>=stealth',shorten >=1pt,auto,node distance=3cm,
    thick,
    main node/.style={circle,draw,font=\sffamily\large\bfseries},
    external node/.style={rectangle,draw,fill=gray!20,font=\sffamily\large\bfseries,minimum size=1cm}]
    
    \node[main node] (IS)          {IS};
    \node[main node] (OS) [below left of=IS]    {OS};
    \node[main node] (PARIS) [below right of=IS] {PARIS};
    \node[external node] (EXT) [below right of=OS]  {External};
    
    \path[every node/.style={font=\sffamily\small}]
    (IS) edge [loop above]         node {25.0\%} (IS)
         edge [bend right=20]      node[left]  {2.7\%}  (OS)
         edge [bend left=20]       node[right] {4.6\%}  (PARIS)
         edge [bend left=15]       node[above right] {0.1\%} (EXT)
    (OS) edge [bend right=20]      node[right] {2.7\%}  (IS)
         edge [loop left]          node {33.7\%} (OS)
         edge                      node[below left] {1.9\%}  (PARIS)
         edge [bend left=15]       node[below] {0.3\%}  (EXT)
    (PARIS) edge [bend right=20]   node[left]  {4.6\%}  (IS)
           edge [bend left=20]     node[right] {1.9\%}  (OS)
           edge [loop right]       node {18.9\%} (PARIS)
           edge [bend left=10]     node[right] {0.1\%} (EXT)
    (EXT) edge [bend left=15]      node[left] {0.3\%} (OS)
         edge [bend right=15]      node[right] {0.2\%} (IS)
         edge [bend right=15]      node[below] {0.2\%} (PARIS)
         edge [loop below]         node {3.1\%} (EXT);
\end{tikzpicture}
\caption{Including external regions}
\label{fig:externalflows}
\end{subfigure}
\caption{Mobility flows between geographical zones in the dataset}
\label{fig:mobility-graphs}
\end{figure}

Fig. ~\ref{fig:interzonalflows} displays inter-zonal flows as a directed graph. Most displacements occur within the same zone, and the flows are fairly balanced between inbound and outbound directions.

% \begin{figure}[!h]
% \begin{center}
		 
% \end{center}
% \caption{Graph of the mobility between areas of Ile-de-France (IS, OS, Paris)}
% \end{figure}

A more detailed inter-departmental flow matrix is shown in Table~\ref{tab:deptflows}, highlighting that intra-department travel dominates, though some cross-department patterns emerge. Although the outer suburbs generate the highest absolute volume of trips, Paris exhibits the highest flow density when normalized by surface area—confirming its central role in the region’s mobility structure.

\begin{table}[H]
\centering
\caption{Percentage of trips between departments}
\label{tab:deptflows}
\begin{tabular}{lcccccccc}
\hline
\textbf{DEP\_D / DEP\_O} & 75 & 77 & 78 & 91 & 92 & 93 & 94 & 95 \\
\hline
75  & 19.7 & 0.5 & 0.5 & 0.4 & 1.9 & 1.5 & 1.5 & 0.5 \\
77  & 0.5  & 8.9 & 0.0 & 0.3 & 0.1 & 0.4 & 0.2 & 0.1 \\
78  & 0.5  & 0.0 & 8.3 & 0.2 & 0.6 & 0.0 & 0.1 & 0.3 \\
91  & 0.4  & 0.3 & 0.2 & 8.9 & 0.3 & 0.0 & 0.3 & 0.0 \\
92  & 1.8  & 0.1 & 0.6 & 0.3 & 8.8 & 0.3 & 0.4 & 0.3 \\
93  & 1.5  & 0.4 & 0.0 & 0.1 & 0.3 & 8.0 & 0.4 & 0.3 \\
94  & 1.4  & 0.2 & 0.1 & 0.3 & 0.5 & 0.4 & 7.0 & 0.1 \\
95  & 0.5  & 0.1 & 0.3 & 0.0 & 0.3 & 0.3 & 0.1 & 7.1 \\
\hline
\end{tabular}
\end{table}

A small fraction of trips extends beyond Île-de-France. These interregional flows are summarized in Fig. ~\ref{fig:externalflows}, where an additional “External” node aggregates all out-of-region destinations.

\subsection{Spatial Dynamics of Mobility Flows}

% Fig.~\ref{fig:trip_records_user} illustrates the distribution of the number of records per individual. Most individuals have, on average, 23 records. To obtain this count, each entry in the \texttt{Date\_EMG} column (where each date represents one record) from the \texttt{trips\_dataset} file is considered. The number of records is calculated per individual, identified by unique values in the \texttt{ID} column. Individuals who do not have a corresponding file in the \texttt{gps\_dataset} folder are excluded from the count, as they have no displacement data available. \jos{Number of records inside the displacement files (including the No\_trips or No\_traces)}

% \begin{figure}[!htb]
%     \centering
%     \includegraphics[width=\textwidth]{Figures/records_count_per_user.png}
%     \caption{Distribution of trip records per individual.}
%     \label{fig:trip_records_user}
% \end{figure}

%\jos{Add a preamble}

\begin{figure}[!htb]
    \centering
    \begin{tiny}
    \subfloat[INSEE map.]{\includegraphics[width=.45\linewidth]{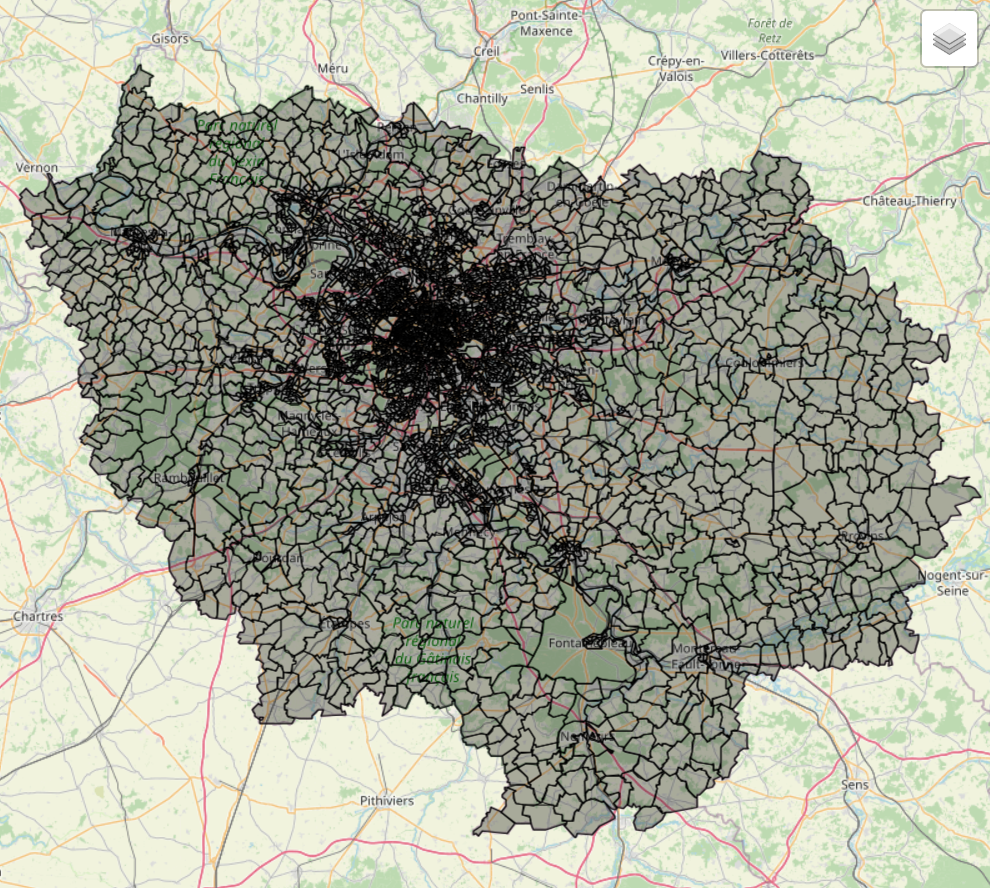}}\hspace*{\fill}
    \subfloat[Visited IRIS map.]{\includegraphics[width=.45\linewidth]{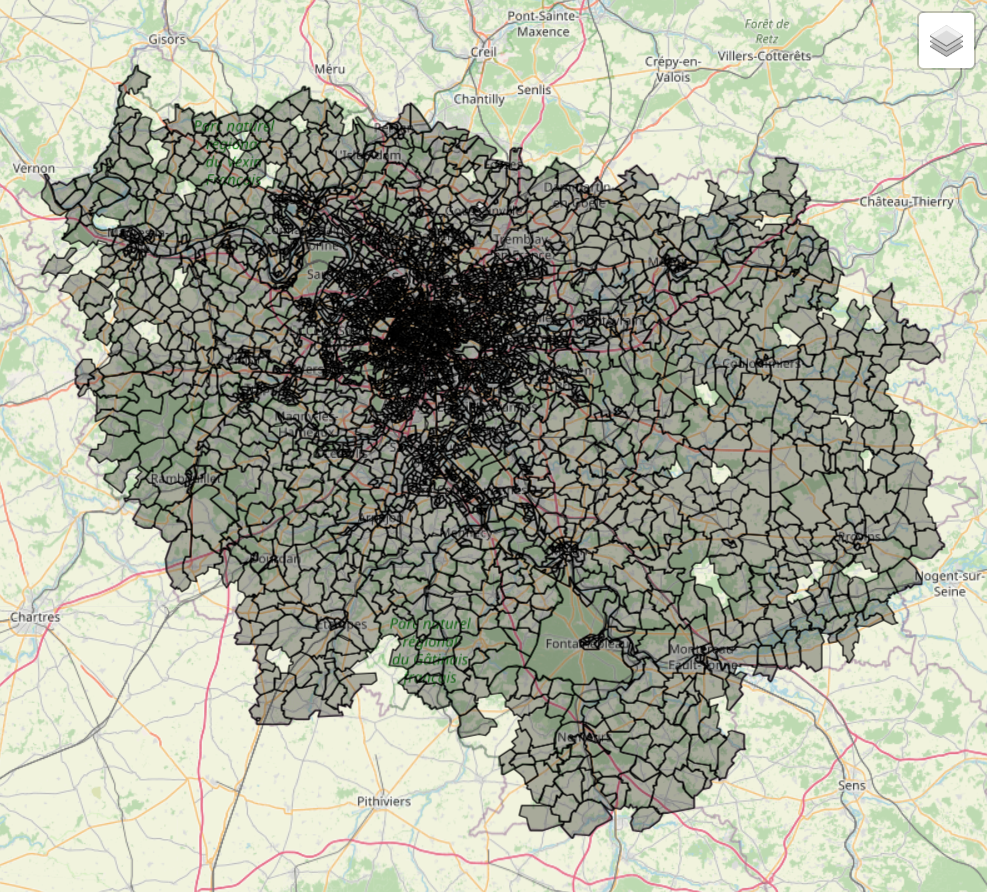}}
    \end{tiny}
    \caption{Paris IRIS organization, according to INSEE's original dataset (5264), and the visited IRIS (5142).}
    \label{fig:paris_iris}
\end{figure}

\paragraph{Observed spatial coverage.}
Fig.~\ref{fig:paris_iris}(a) displays the complete set of 5,264 IRIS units in the Île-de-France region, as defined by INSEE (Institut national de la statistique et des études économiques). Fig.~\ref{fig:paris_iris}(b) shows the subset of 5,142 IRIS units that were actually visited at least once by an individual, based on the mobility traces available in the \texttt{gps\_dataset}.

\paragraph{Average number of trips per IRIS.}

\begin{figure}[H]
    \centering
    \begin{tiny}
    \subfloat[Per date.]{\includegraphics[width=.49\linewidth]{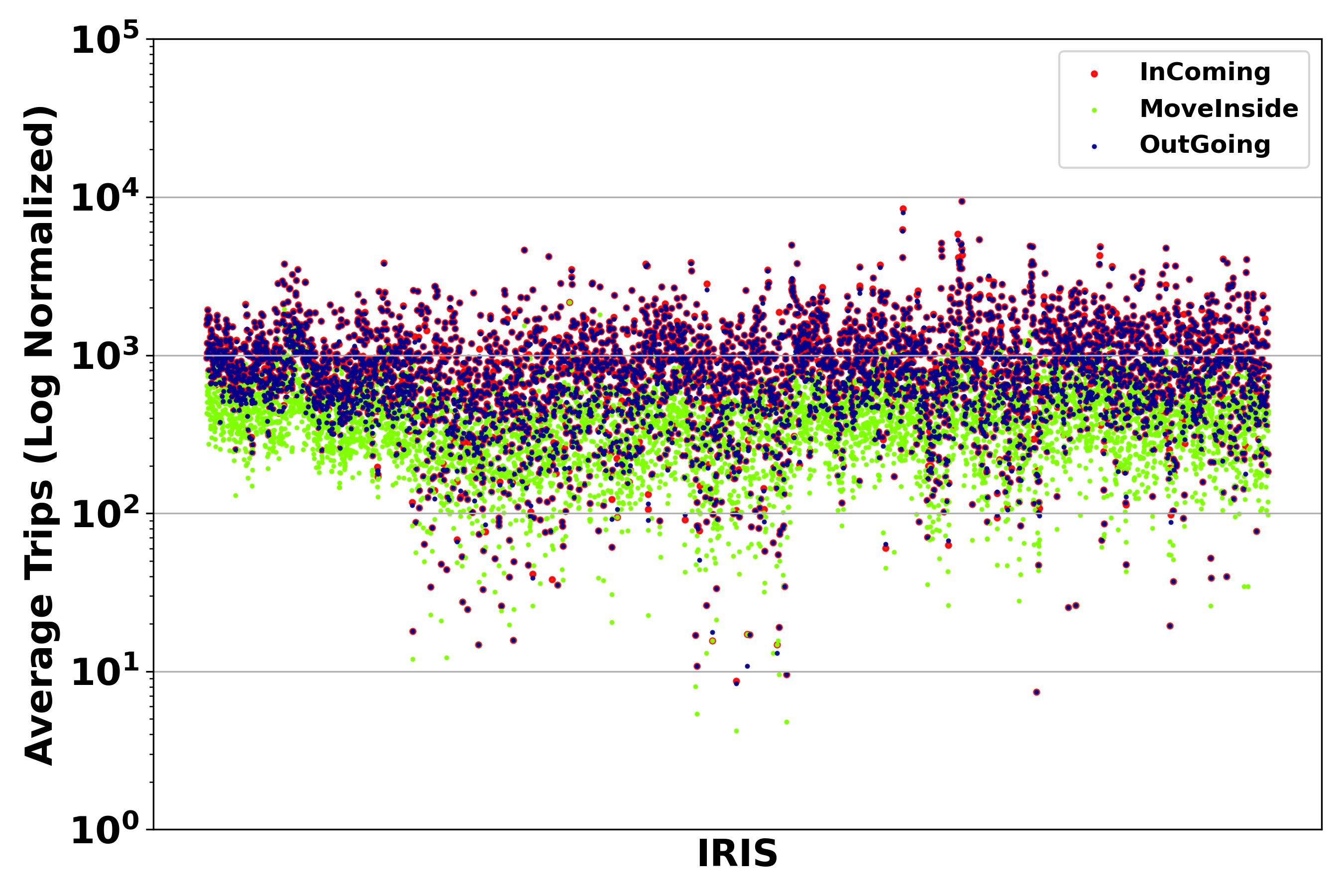}}\hspace*{\fill}
    \subfloat[Per hour.]{\includegraphics[width=.49\linewidth]{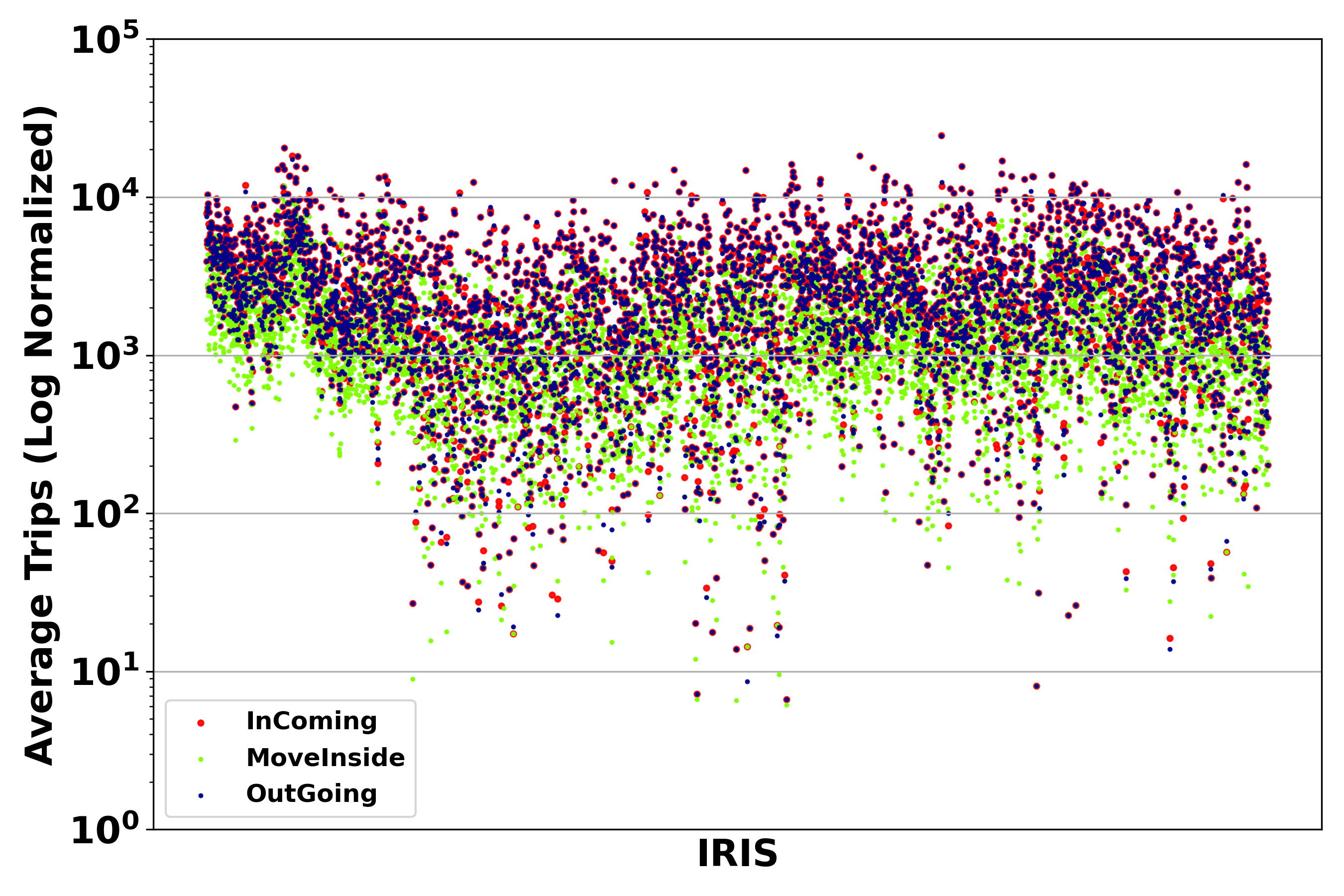}}
    \end{tiny}
    \caption{The average number of MoveInside, InComing, and OutGoing trips in each IRIS, per date and per hour.}
    \label{fig:in_outs}
\end{figure}

Fig.~\ref{fig:in_outs} presents the average number of trips per IRIS unit, computed separately for each hour and each day, and categorized into three types: \textit{MoveInside}, \textit{InComing}, and \textit{OutGoing}. These categories refer respectively to trips occurring entirely within an IRIS, trips entering the IRIS from another location, and trips exiting the IRIS toward another location. The mapping of each IRIS unit to a unique index on the x-axis follows a fixed departmental ordering as detailed in Table~\ref{tab:iris_mapping}, ensuring reproducibility and interpretability of visualized spatial patterns.

We obtain these metrics by processing the sequences of positions $P = \{p_1, p_2, ..., p_n\}$ recorded in the \texttt{gps\_dataset} for each trip. If, on a given date $d$, an individual is at location $p_1$ in IRIS$_1$ at time $t_1$, and later at $p_2$ in IRIS$_2$ at time $t_2$, then $p_1$ constitutes an exit from IRIS$_1$ and $p_2$ an entry into IRIS$_2$. Each such event is recorded along with the corresponding date, time, and the associated trip weight (\texttt{Weight\_Day}) provided in the \texttt{trips\_dataset}. The \textit{MoveInside} category accounts for successive positions $p_i$, $p_{i+1}$ that fall within the same IRIS.

Trip counts for each IRIS and time interval are obtained by summing these weights. We then compute average values across the entire observation period for both daily and hourly resolutions.

As shown in Fig.~\ref{fig:in_outs}(a), the average number of \textit{InComing} and \textit{OutGoing} trips per day is remarkably similar across IRIS units, and both are significantly higher than \textit{MoveInside} trips. This suggests that most IRIS serve primarily as origins or destinations rather than internal circulation zones. Such behavior is typical of areas with high transient traffic, such as transport hubs, business districts, or commercial zones.

Fig.~\ref{fig:in_outs}(b) further highlights hourly variations. Several IRIS exhibit strong temporal dynamics with pronounced entry and exit volumes, and a wider variance in all three flow categories. These observations point to heterogeneity in population density and infrastructure availability across the region, particularly in relation to public transit and daily commuting behaviors.

\begin{table*}[!htb]
    \centering
    \caption{Mapping of IRIS codes to unique integer values.}
    \label{tab:iris_mapping}
    \begin{tabularx}{\textwidth}{@{}lX lX@{}}
        \toprule
        \textbf{Department} & \textbf{Mapping Range} & \textbf{Department} & \textbf{Mapping Range} \\
        \midrule
        75 & from 0 to 991 & 92 & from 2839 to 3454 \\
        77 & from 992 to 1698 & 93 & from 3455 to 4067 \\
        78 & from 1699 to 2341 & 94 & from 4068 to 4597 \\
        91 & from 2342 to 2838 & 95 & from 4598 to 5141 \\
        \bottomrule
    \end{tabularx}
\end{table*}

% Fig.~\ref{fig:trips_total_iris} presents the estimated number of trips in each IRIS. A trip is said to pass through an IRIS if at least one of its positions can be located within the IRIS. This verification is performed using the locations of the trips in \texttt{gps\_dataset}. Subsequently, the sum of the weights of each trip, provided by the \texttt{trips\_dataset}), is computed per date and per hour.

\paragraph{Spatio-temporal distribution of trip counts.}

\begin{figure}[H]
    \centering
    \begin{tiny}
    \subfloat[Per date]{\includegraphics[width=.49\linewidth]{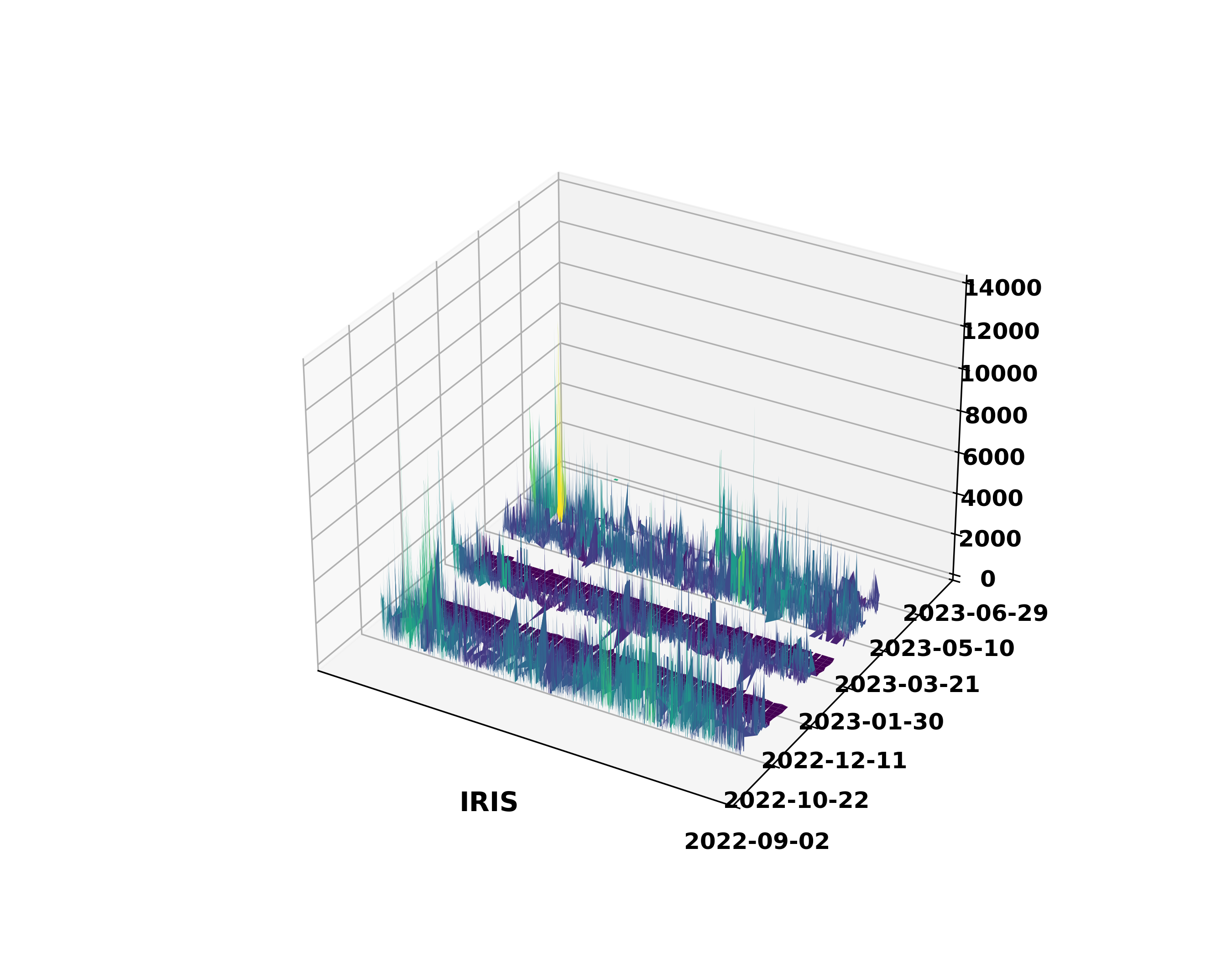}}\hspace*{\fill}
    \subfloat[Per hour]{\includegraphics[width=.49\linewidth]{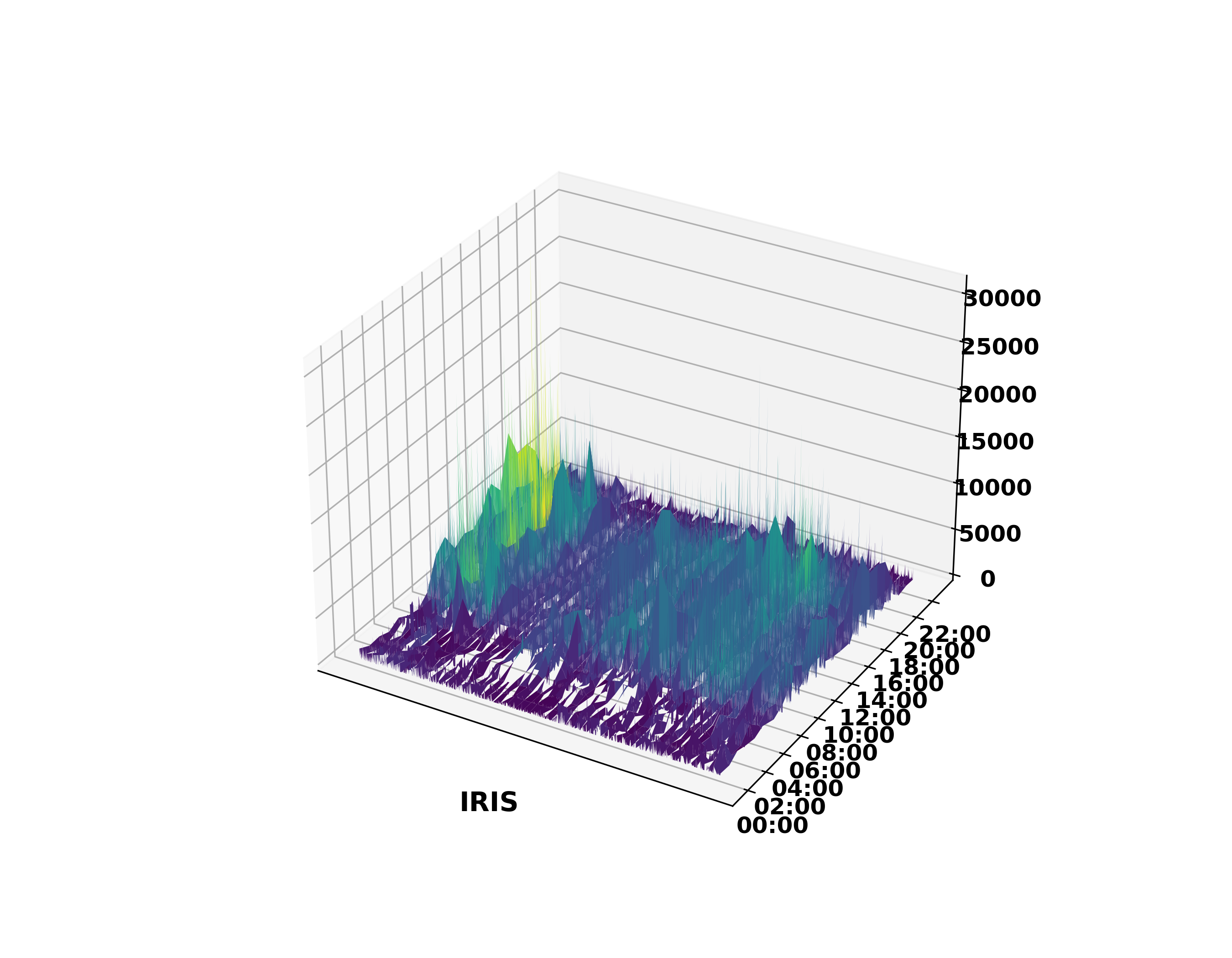}}
    \end{tiny}
    \caption{Number of trips passing by each IRIS per date and per hour.}
    \label{fig:trips_total_iris}
\end{figure}

Fig.~\ref{fig:trips_total_iris} presents the estimated number of trips observed within each IRIS over time. A trip is considered to pass through a given IRIS if at least one of its recorded GPS positions falls within the geographical boundaries of that IRIS. This spatial association is established using the location data provided in the \texttt{gps\_dataset}. For each identified IRIS crossing, the corresponding trip weight from the \texttt{trips\_dataset} (specifically the \texttt{Weight\_Day} value) is used to compute the total trip volume, aggregated by date and hour.

In Fig.~\ref{fig:trips_total_iris}(a), which depicts daily totals, we observe substantial disparities across IRIS units. Some IRIS exhibit significantly higher trip counts, with one reaching up to 8{,}000 trips in a single day. These peaks likely correspond to key transport interchanges or densely trafficked urban locations. The presence of white horizontal bands in the heatmap suggests the absence of trip data for certain days, which may result from device inactivity, data collection interruptions, or filtering during preprocessing.

Fig.~\ref{fig:trips_total_iris}(b) illustrates the distribution of trips per hour across IRIS units. Similar spatial patterns emerge, with a subset of IRIS consistently showing elevated trip volumes throughout the day. Trip activity is minimal during the early morning hours (00:00 to 06:00), with a marked increase beginning around 06:00, reflecting typical urban mobility cycles driven by commuting patterns.

% \paragraph{3D View} Fig.~\ref{fig:trips_total_iris} presents the estimated number of trips in each IRIS. A trip is considered to pass through an IRIS if at least one of its recorded positions falls within the boundaries of that IRIS. This verification is performed using the location data from the \texttt{gps\_dataset}. Subsequently, the sum of the corresponding trip weights (provided in the \texttt{trips\_dataset}) is computed and aggregated by date and hour.

% In Fig.~\ref{fig:trips_total_iris}(a), we observe that some IRIS have a high number of trips compared to others, one IRIS reaches approximately 8,000 trips in a single day. This concentration suggests the presence of major transit hubs or highly frequented areas. The presence of white bands across the timeline may be caused by sequences of days with no recorded trip data, data collection gaps, device inactivity, or filtering during preprocessing. Fig.~\ref{fig:trips_total_iris}(b) illustrates the estimated number of trips per hour in each IRIS. Once again, certain IRIS stand out with consistently high hourly trip volumes, reinforcing the idea of their central role in urban mobility. Trip volumes are significantly lower during nighttime hours (00:00 to 06:00), and begin to rise in the early morning.

\begin{figure}[H]
    \centering
    \begin{tiny}
    \subfloat[Per date]{\includegraphics[width=.49\linewidth]{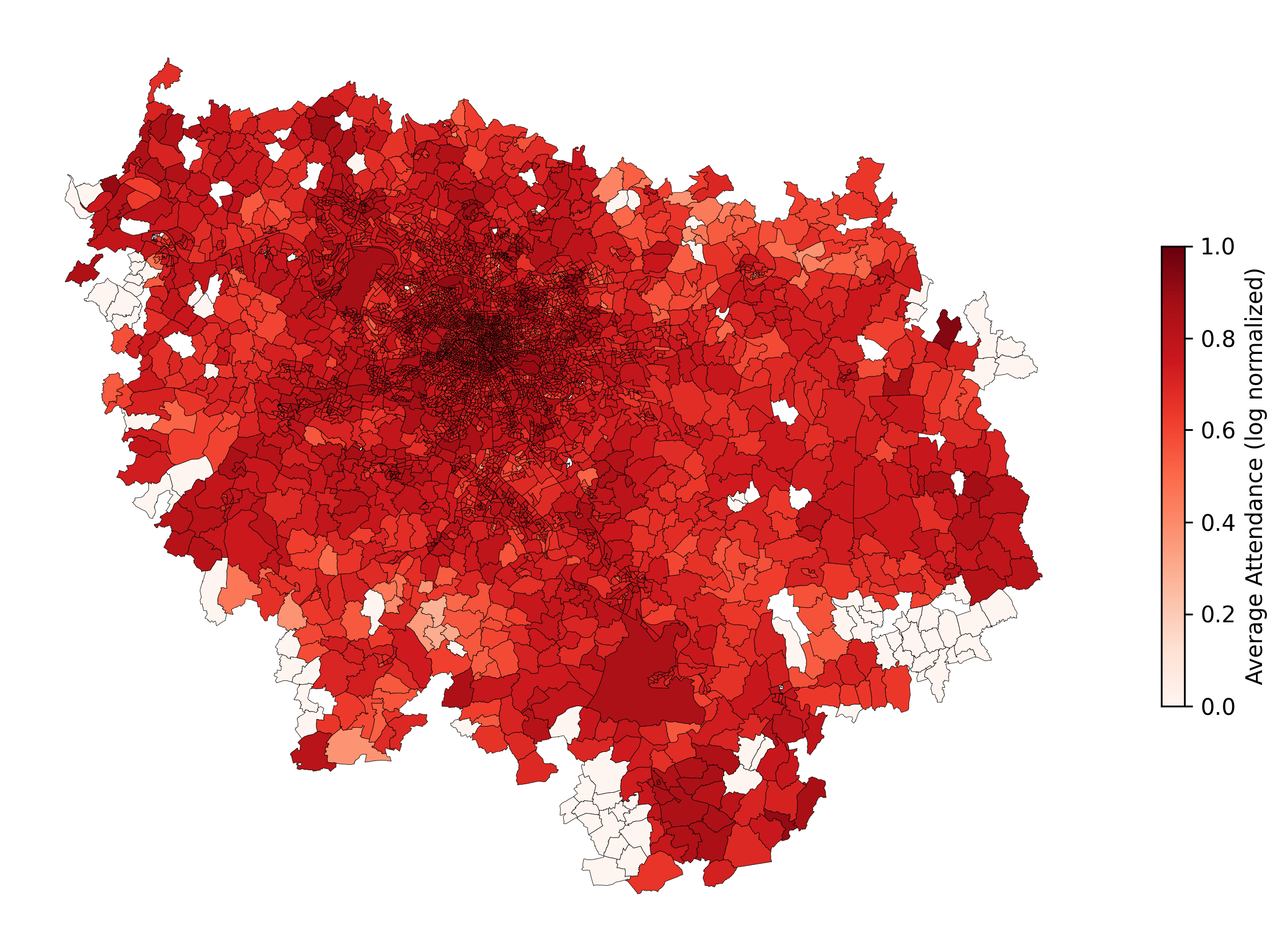}}\hspace*{\fill}
    \subfloat[Per hour]{\includegraphics[width=.49\linewidth]{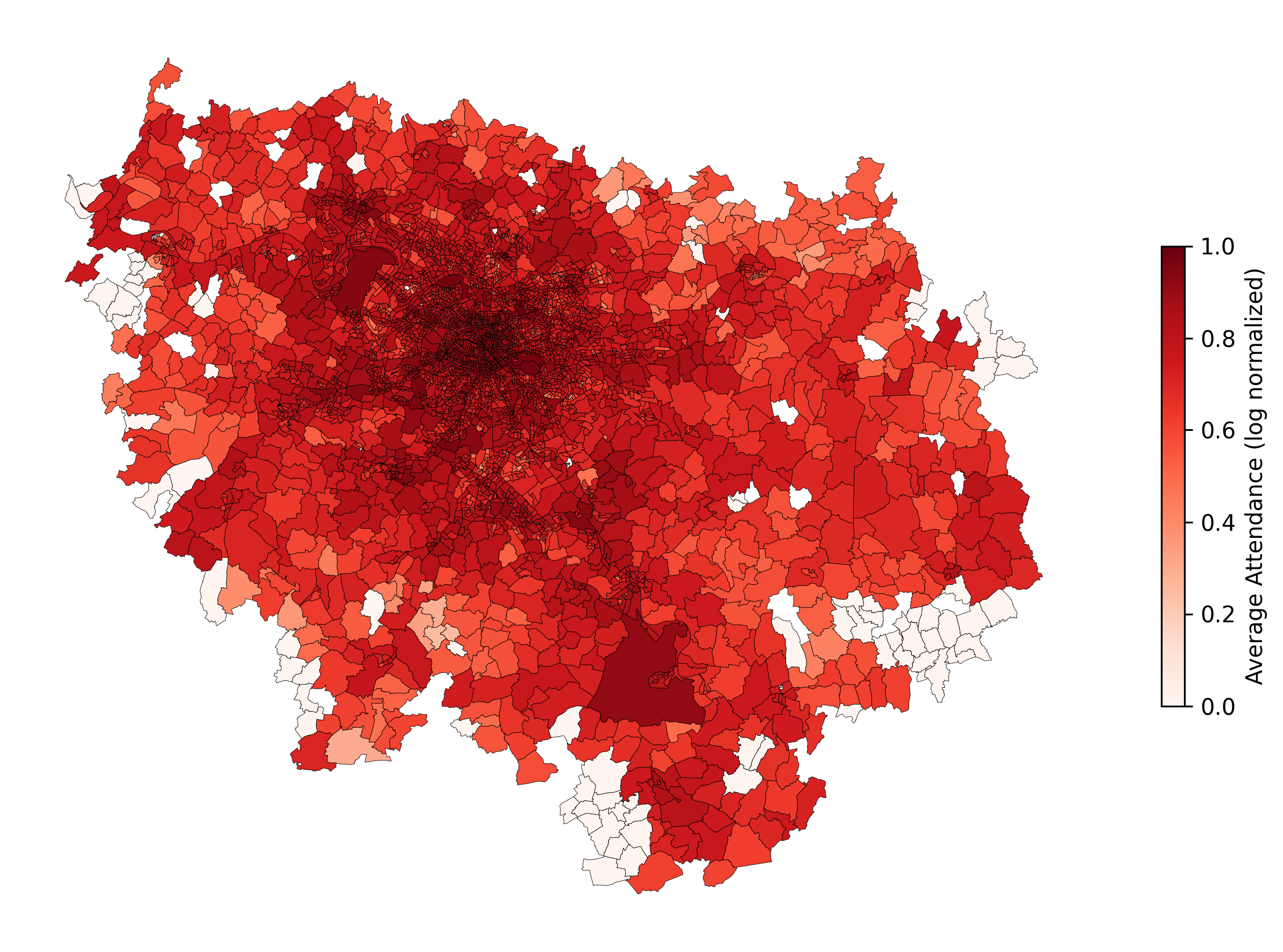}}
    \end{tiny}
    \caption{Spatial distribution of attendance in each IRIS per date and per hour.}
    \label{fig:iris_attendance}
\end{figure}

\paragraph{Attendance and density metrics.}
Fig.~\ref{fig:iris_attendance} illustrates the spatial distribution of average attendance in each IRIS, aggregated by date and by hour. The attendance at a given IRIS $s$ and time $t$ is defined as $a_s^t = \text{MoveInside} + \frac{(\text{InComing} + \text{OutGoing})}{2}$. Comparing the average attendance across Fig.~\ref{fig:iris_attendance}(a)--(b) reveals similar spatial patterns across daily and hourly aggregations. However, higher attendance levels are observed in IRIS units located in the central part of the Île-de-France region, particularly within Paris. This observation is consistent with expectations for a metropolitan core, where the concentration of workplaces, public services, commercial districts, and transport infrastructure contributes to consistently elevated mobility and presence.

% \paragraph{Attendance and Density} Fig.~\ref{fig:iris_attendance} illustrates the spatial distribution of average attendance in each IRIS, aggregated by date and by hour. The attendance at a given IRIS $s$ at time $t$ is defined as $a_s^t = MoveInside + \frac{(InComing + OutGoing)}{2}$.  When comparing the average attendance in Figures~\ref{fig:iris_attendance}(a)-(b), we observe similar spatial patterns. But an increase in attendance is noticeable in IRIS units located in the central part of the Île-de-France region, particularly in Paris. Which is aligned with expectations for a metropolitan core, where a concentration of workplaces, public services, commercial areas, and transport infrastructure contributes to consistently higher mobility and presence.

\begin{figure}[!htb]
    \centering
    \begin{tiny}
    \subfloat[Per date.]{\includegraphics[width=.49\linewidth]{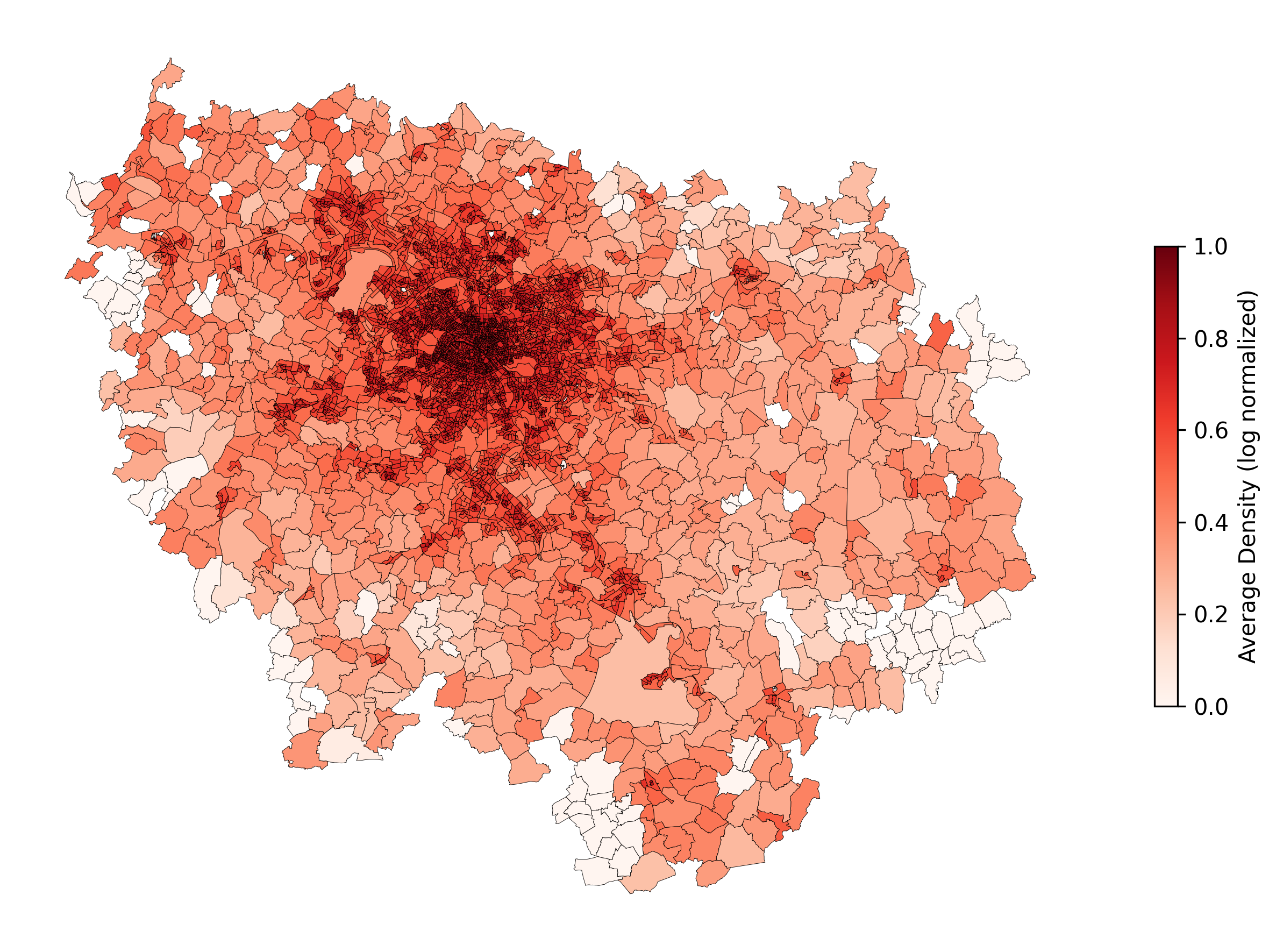}}\hspace*{\fill}
    \subfloat[Per hour.]{\includegraphics[width=.49\linewidth]{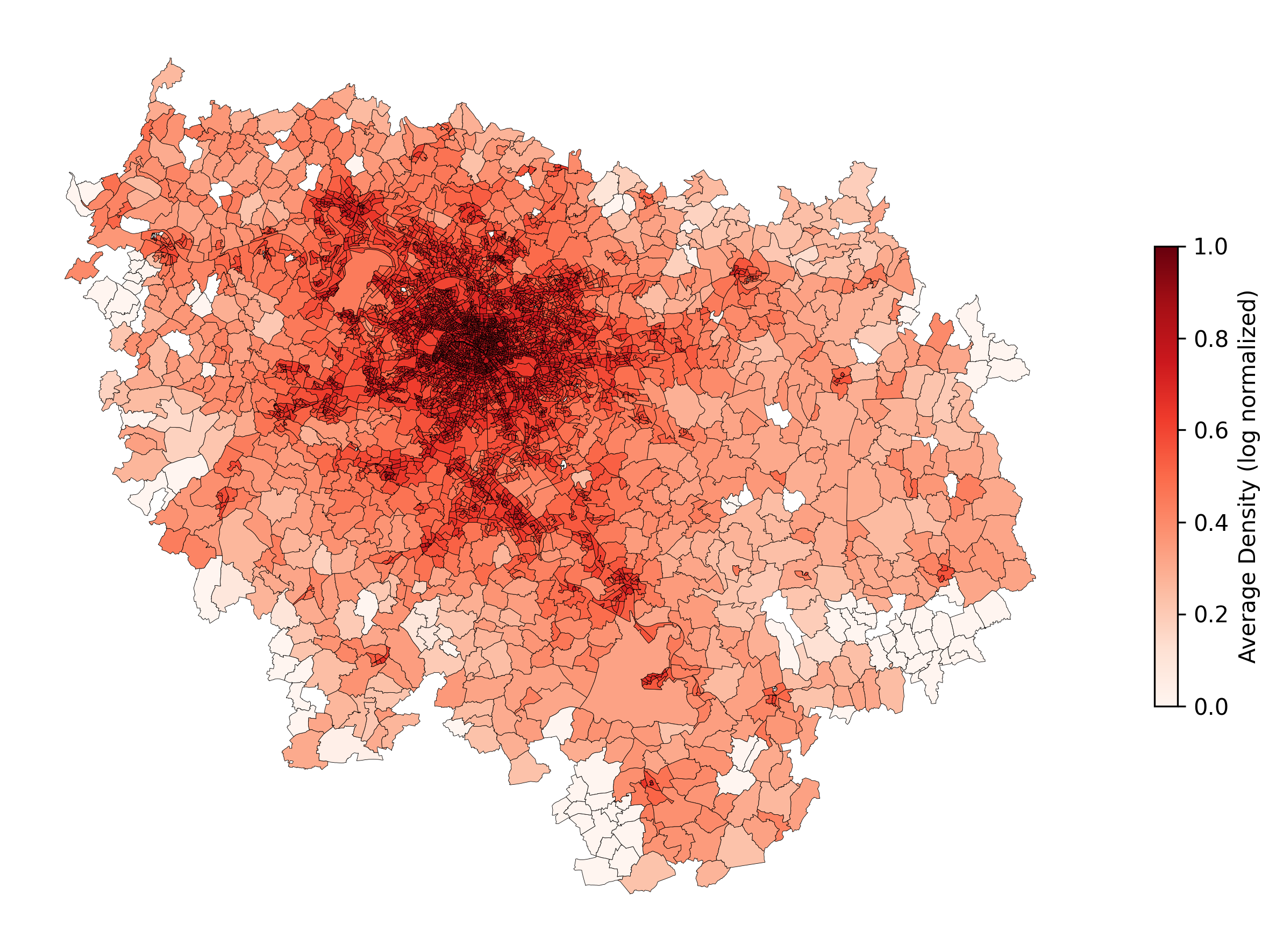}}
    \end{tiny}
    \caption{Spatial distribution of density (Attendance/IRIS Area) per date and per hour.}
    \label{fig:trips_density}
\end{figure}

Fig.~\ref{fig:trips_density} shows the spatial distribution of average population density per IRIS, computed at both daily and hourly resolutions. The density for IRIS $s$ at time $t$ is defined as $D_s^t = \frac{a_s^t}{A_s}$, where $a_s^t$ is the attendance (as defined above), and $A_s$ is the surface area of IRIS $s$ in square kilometers. Similar to the attendance patterns, the average density values exhibit stable spatial trends across temporal resolutions. The highest densities are concentrated in central Paris, reflecting intense population activity. Notably, some IRIS units located in more peripheral areas of Île-de-France also show elevated density levels. These may correspond to compact urban cores, strategic transport hubs, or suburban centers with high localized activity despite their distance from the metropolitan center.

% Fig.~\ref{fig:trips_density} shows the spatial distribution of average population density in each IRIS, calculated per day and per hour. The density at IRIS $s$ at time $t$ is defined as $D_s^t = \frac{a_s^t}{Area_s}$, where $a_s^t$ represents the attendance at that time (as previously defined), and $A_s$ is the area of IRIS $s$, measured in square kilometers. Similar to attendance, the average density values exhibit similar values when comparing daily and hourly time intervals. The highest densities are concentrated in central Paris, reflecting the region's high population activity. It is also noteworthy that some IRIS units located farther from the center of Île-de-France show elevated density levels. These may correspond to compact urban centers, key transport nodes, or suburban hubs with concentrated activity despite their geographic distance.

\paragraph{Trip length distributions.}

\begin{figure}[!htb]
    \centering
    \includegraphics[width=0.5\textwidth]{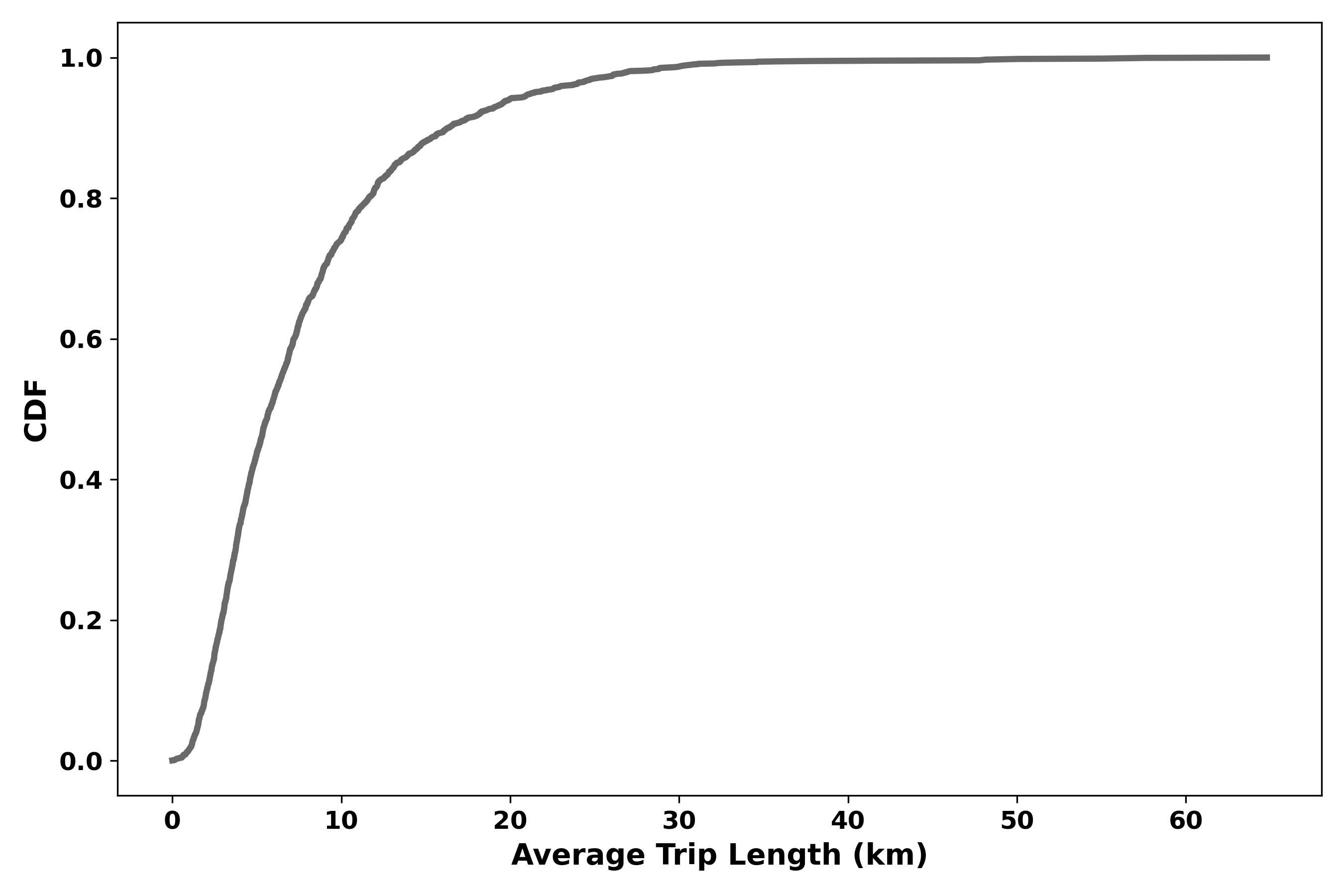}
    \caption{Trips' average length per user.}
    \label{fig:trips_avg_length}
\end{figure}

Fig.~\ref{fig:trips_avg_length} presents the empirical cumulative distribution function (CDF) of the average trip length per individual. Approximately 90\% of individuals travel an average distance of 20 km or less per trip. Trip lengths are computed by summing the distances between consecutive recorded positions $P = \{p_1, p_2, ..., p_n\}$ for each individual, using data from the \texttt{gps\_dataset}. The start and end timestamps defining the boundaries of each trip are retrieved from the columns \texttt{Date\_O}, \texttt{Time\_O}, \texttt{Date\_D}, and \texttt{Time\_D} in the \texttt{trips\_dataset}.

% \paragraph{Trips lengths}
% Fig.~\ref{fig:trips_avg_length} presents the cumulative distribution of the average trip length per individual. Approximately 90\% of individuals travel an average distance of 20 km or less per trip. Trip lengths are calculated by summing the distances between consecutive positions $P = \{p_1, p_2, ..., p_n\}$ for each individual, using the data from the \texttt{gps\_dataset}. To identify the set of positions belonging to a specific trip, we use the timestamps provided in the \texttt{trips\_dataset}. The columns \texttt{Date\_O}, \texttt{Time\_O}, \texttt{Date\_D}, and \texttt{Time\_D} indicate the start and end date and time of each trip.

\begin{figure}[H]
    \centering
    \subfloat[\scriptsize Average trip lengths per weekday. Each point represents an individual’s average trip length aggregated over all trips made on that weekday.]
    {\includegraphics[width=.49\linewidth]{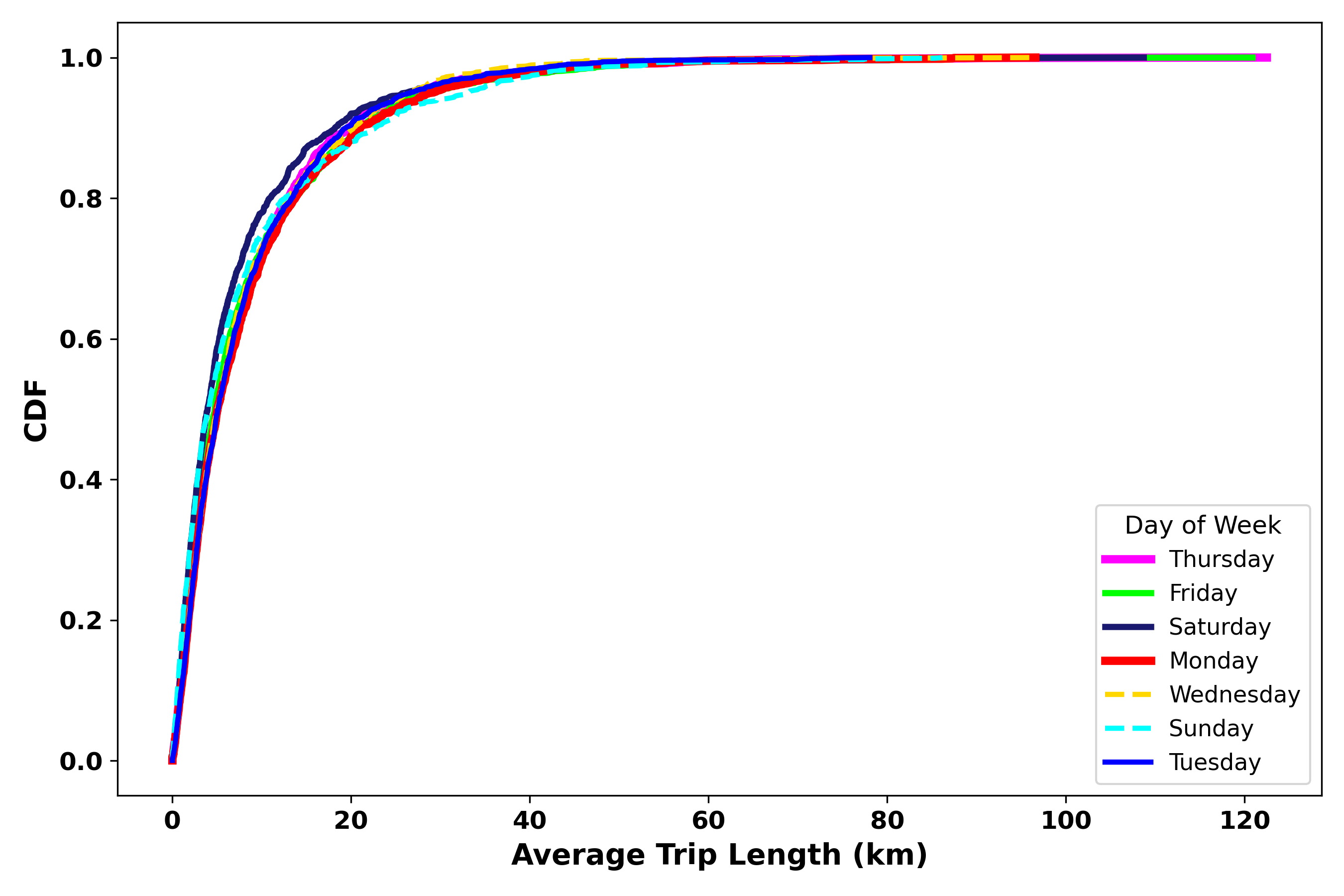}}\hspace*{\fill}
    \subfloat[\scriptsize Average trip lengths per individual per time period of the day, aggregated over all days.]
    {\includegraphics[width=.49\linewidth]{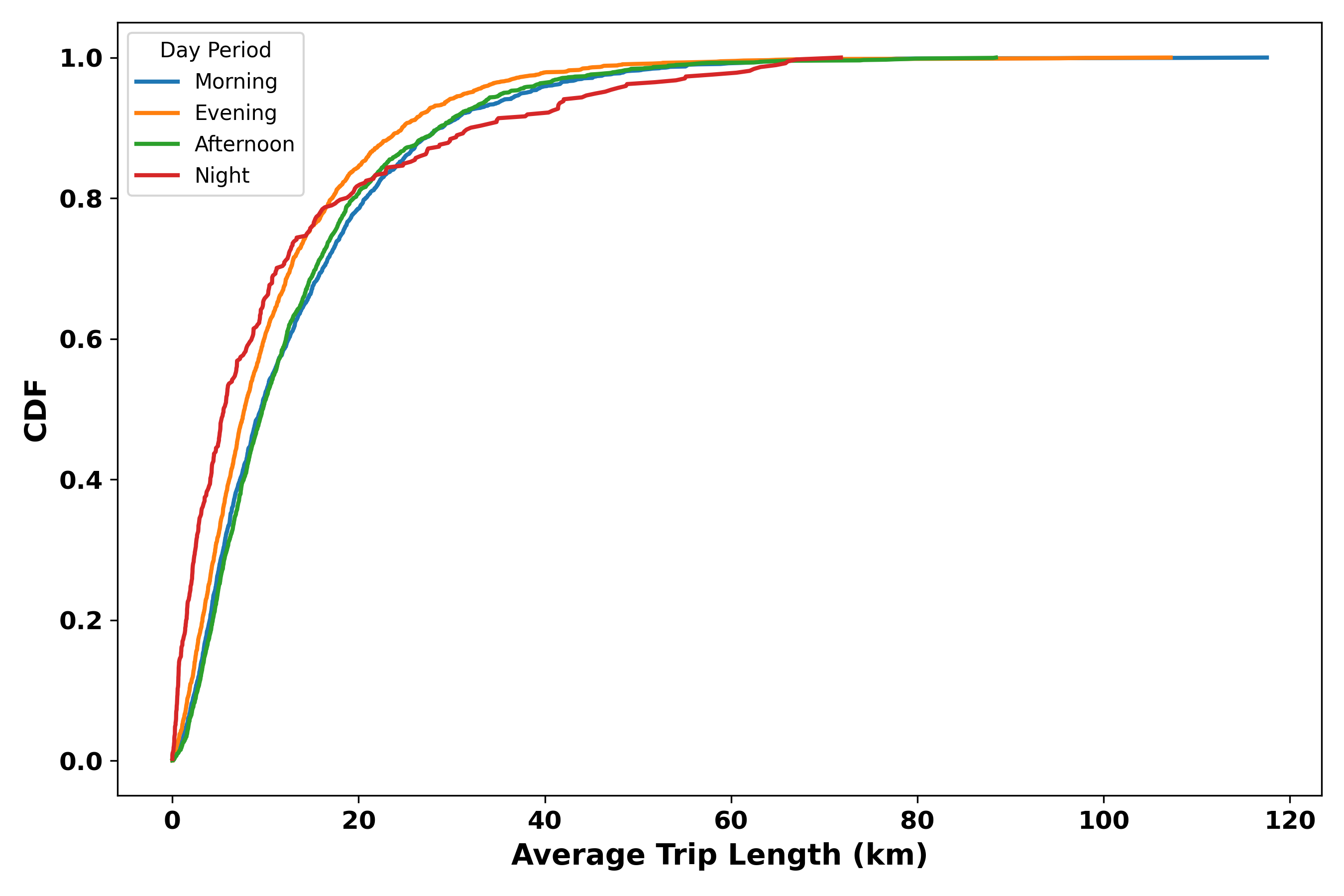}}
    \caption{Empirical cumulative distribution of average trip lengths per individual, across weekdays (left) and time-of-day periods (right).}
    \label{fig:trips_avg_length2}
\end{figure}

Fig.~\ref{fig:trips_avg_length2}(a) displays the distribution of average trip lengths computed per individual for each weekday. Each point in the distribution represents one individual's average trip length over all trips made on that weekday, without distinguishing trips across specific time slots. The longest trips are observed on Thursdays, followed by Fridays, Saturdays, and Mondays. Tuesdays exhibit the shortest average trip lengths.

In Fig.~\ref{fig:trips_avg_length2}(b), the analysis is stratified by four time periods: Morning (06:00–11:59), Afternoon (12:00–17:59), Evening (18:00–23:59), and Night (00:00–05:59). For each individual, an average trip length is computed for each period across all days. The cumulative distribution curves reveal that longer trips are more likely during morning and evening hours, likely reflecting commuting behavior. Shorter trips are more frequent during the night and afternoon periods, with night-time trips being the shortest on average.

% From Fig.~\ref{fig:trips_avg_length2}(a), we observe that the longest average trips occur on Thursdays, followed by Fridays, Saturdays, Mondays, Wednesdays, Sundays, and finally Tuesdays, which show the shortest average trip lengths. On all days of the week, over 80@\% of individuals travel an average distance of 20 km or less per trip. This suggests that most mobility behavior remains local or regional in scale, with only a small portion of the population engaging in longer-distance travel on a regular basis.

% For the distributions in Fig.~\ref{fig:trips_avg_length2}(b), the day is divided into four periods: Morning (06:00–11:59), Afternoon (12:00–17:59), Evening (18:00–23:59), and Night (00:00–05:59). The longest trips tend to occur during the morning and evening periods, the cumulative distribution function (CDF) curve rises more rapidly during the night and afternoon periods, indicating that shorter trips are more common during these times. In particular, trips at night are the shortest on average.

% \begin{figure}[!htb]
%     \centering
%     \begin{tiny}
%     \subfloat[Average trip length per weekday. \jos{No distinction of the individual}]{\includegraphics[width=.49\linewidth]{Figures/trips_avg_length_per_user_per_weekday.png}}\hspace*{\fill}
%     \subfloat[Average trip length per user per day period. \jos{Per user in Morning regardless of the day I have one aggreageted value... You plot the distrib where every item a user's value.}]{\includegraphics[width=.49\linewidth]{Figures/trips_avg_length_per_day_period.png}}
%     \end{tiny}
%     \caption{The average trip length per user for different weekday, and for different periods of the day.}
%     \label{fig:trips_avg_length2}
% \end{figure}

\begin{figure}[!htb]
    \centering
    \begin{tiny}
    \subfloat[Per date.]{\includegraphics[width=.49\linewidth]{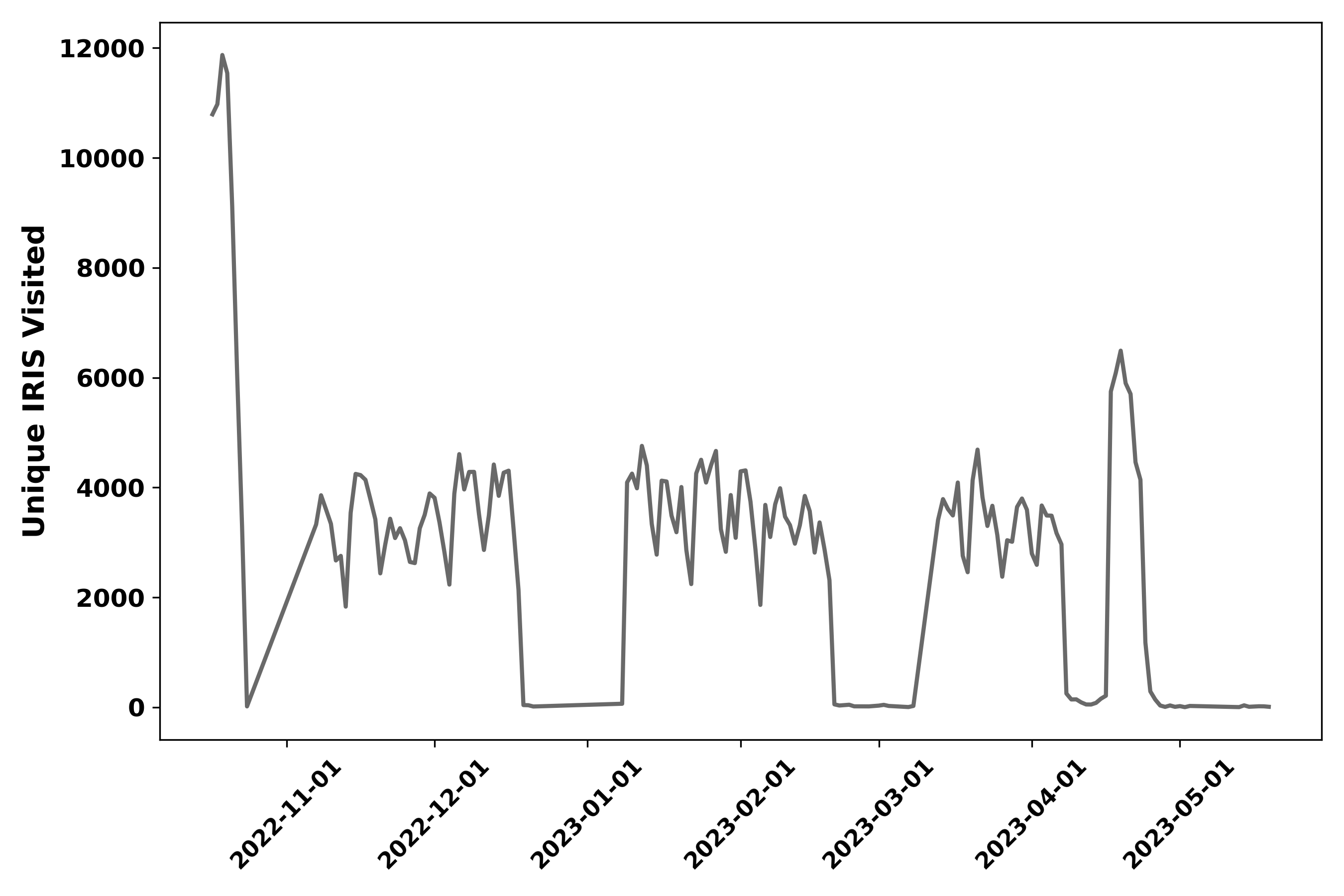}}\hspace*{\fill}
    \subfloat[Per hour.]{\includegraphics[width=.49\linewidth]{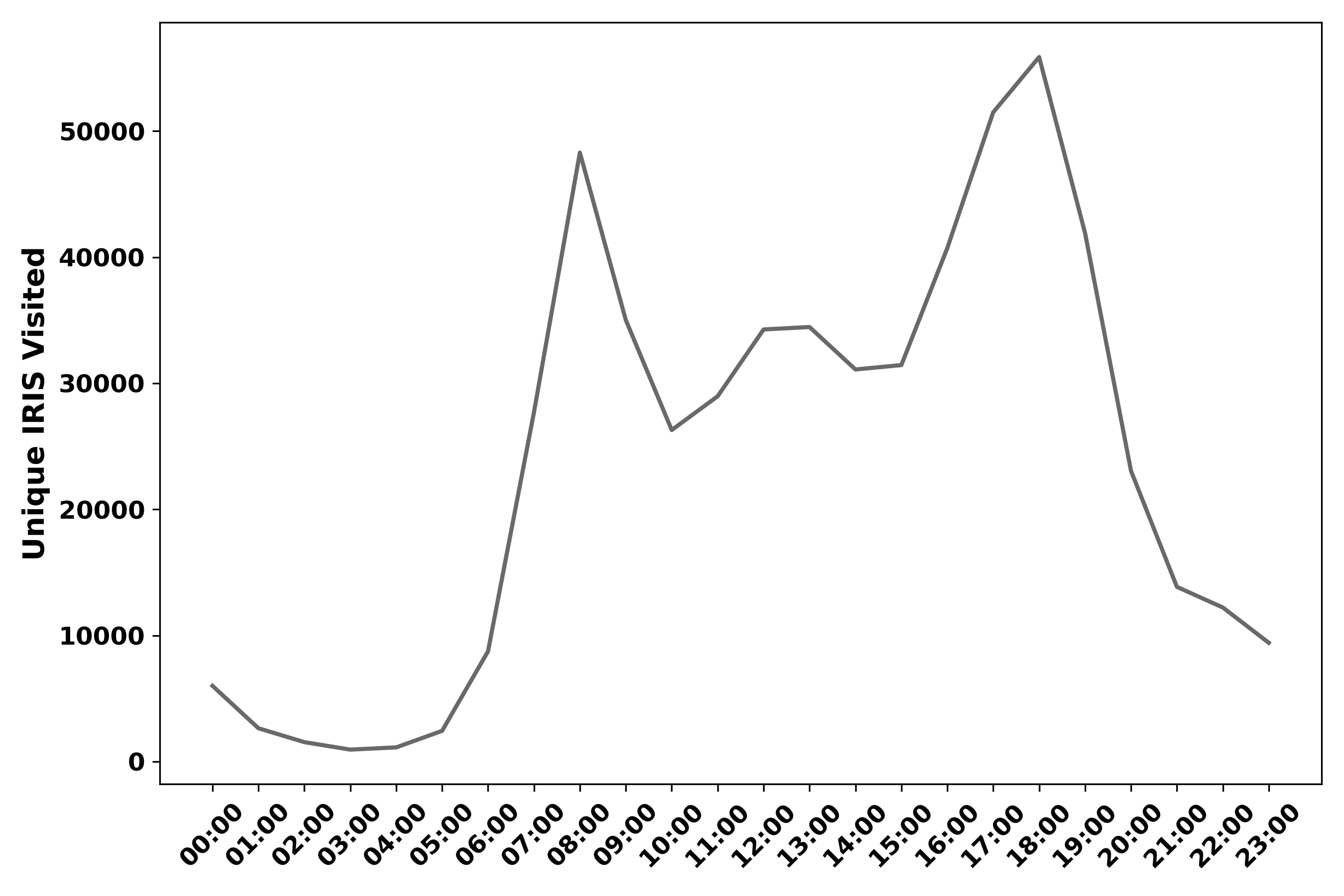}}
    \end{tiny}
    \caption{Number of unique visited locations (IRIS) per date and per hour.}
    \label{fig:unique_iris_visits_user}
\end{figure}

% \paragraph{Uniquely visited IRIS per time interval.}
% Fig.~\ref{fig:unique_iris_visits_user} shows the number of unique IRIS visited by individuals, aggregated by date and by hour. Visited IRIS are identified using the location data from the \texttt{gps\_dataset} for each individual. If a location $p$, recorded at timestamp $t$, falls within the boundaries of an IRIS $s$, that IRIS is registered as a visited location. The results shown in Figures~\ref{fig:unique_iris_visits_user}(a) and~(b) are obtained by grouping the unique IRIS visited by individuals on a daily and hourly basis, respectively.
\paragraph{Unique IRIS units visited per time interval.}
Fig.\ref{fig:unique_iris_visits_user} presents the number of unique IRIS units visited by the all individuals, aggregated by date and by hour. An IRIS unit is considered visited if at least one recorded position $p$, captured at timestamp $t$, falls within its boundaries. These locations are identified using the spatial coordinates from the \texttt{gps\_dataset} for each individual. The figures show the number of distinct IRIS units visited per individual per date (Fig.\ref{fig:unique_iris_visits_user}(a)) and per hour (Fig.~\ref{fig:unique_iris_visits_user}(b)).

\begin{figure}[H]
    \centering
    \begin{tiny}
    \subfloat[Number of unique visited per day period.]{\includegraphics[width=.49\linewidth]{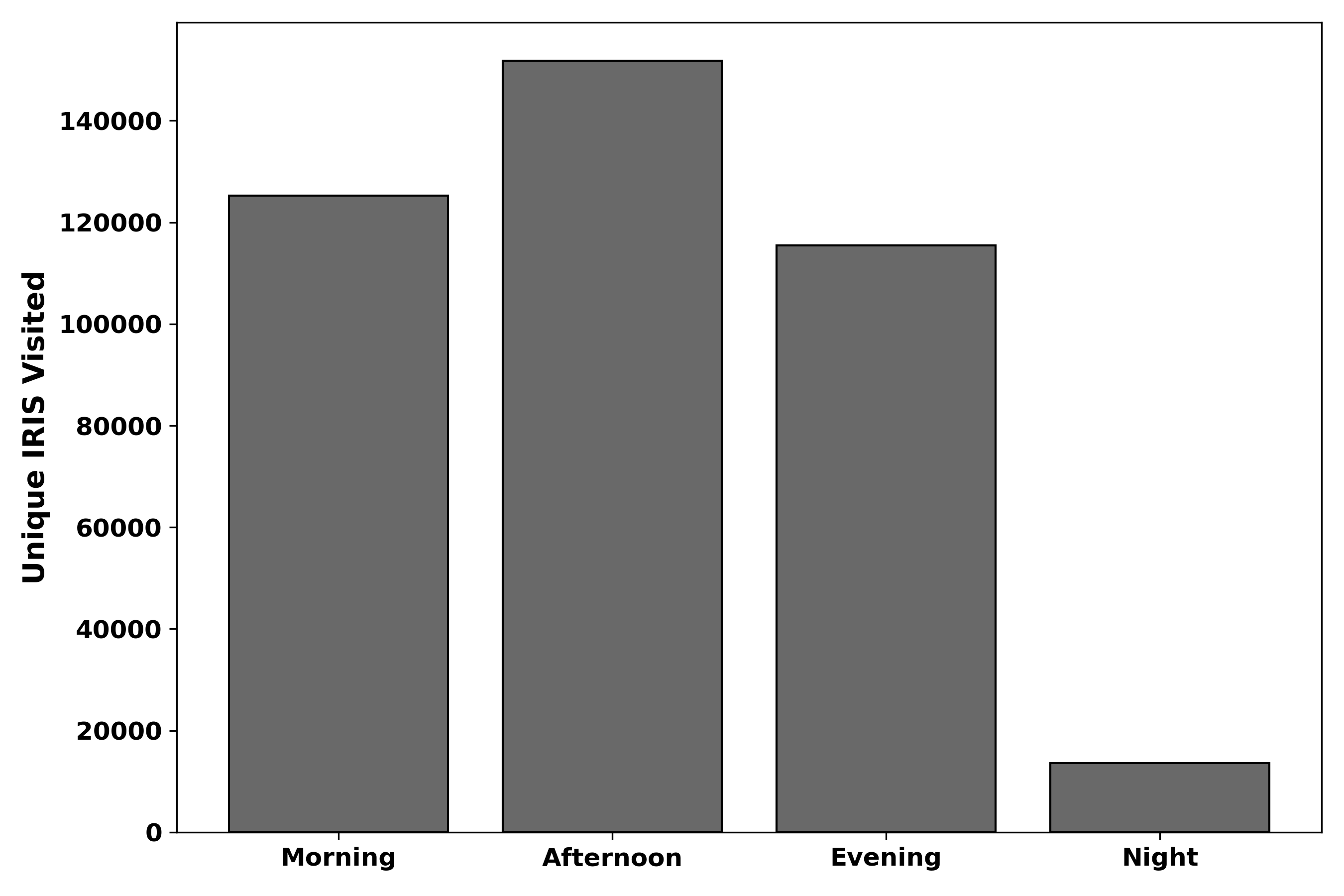}}
    \subfloat[Distribution of unique visits per user.]{\includegraphics[width=.49\linewidth]{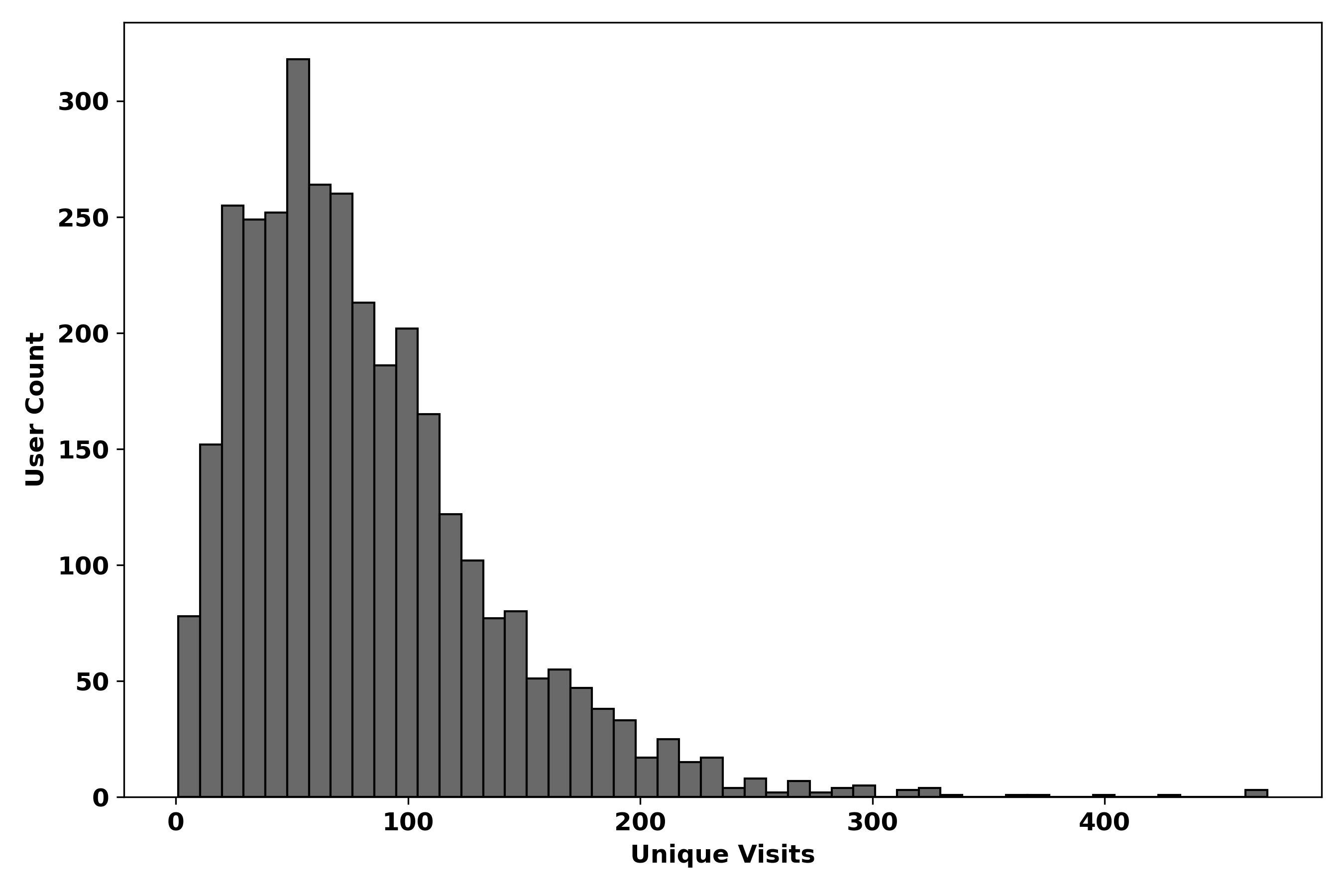}}\hspace*{\fill}
    \end{tiny}
    \caption{Number of unique visited IRIS per day period and Per-user distribution.}
    \label{fig:unique_iris_visits2}
\end{figure}

In Fig.\ref{fig:unique_iris_visits2}(a), the number of unique IRIS units visited is shown for different time periods throughout the day. The afternoon period exhibits the highest number of unique visits, likely reflecting increased activity during working hours and afternoon errands. In contrast, the lowest number of unique visits occurs during the night, consistent with reduced mobility and limited activity during late-night hours. Finally, Fig.\ref{fig:unique_iris_visits2}(b) illustrates the distribution of the total number of unique IRIS units visited per individual over the observation period. Some individuals are observed to have visited up to 81 distinct IRIS units, indicating a high level of spatial mobility.

% Fig.~\ref{fig:unique_iris_visits2}(a) presents the distribution of unique IRIS units visited per individual, with individuals visiting up to 81 unique IRIS locations on average. Fig.~\ref{fig:unique_iris_visits2}(b) shows the number of unique IRIS visited during different periods of the day. The results indicate that the highest number of visits occurs in the afternoon. The lowest number of unique visits is recorded during the night, reflecting the expected reduction in mobility during late hours when most individuals are at home and fewer services or destinations are active.

\subsection{Transport Mode and Temporal Analysis}
Now we explore how individuals move throughout the week by analyzing the main transportation modes, multimodal trip chains, and spatiotemporal flows of departures and arrivals. We examine both the frequency of use per mode and how trips are distributed across time windows and the Île-de-France region.

\begin{figure}[H]
\centering

\begin{subfigure}[t]{0.32\linewidth}
  \centering
  \includegraphics[width=\linewidth]{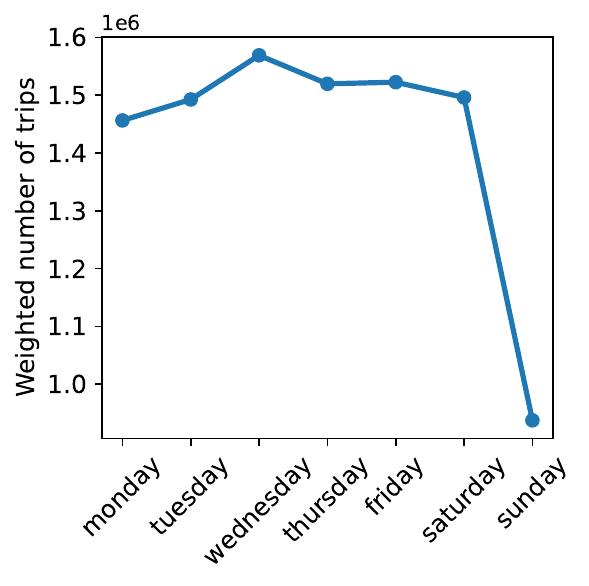}
  \caption{Private car (driver)}
\end{subfigure}
\begin{subfigure}[t]{0.32\linewidth}
  \centering
  \includegraphics[width=\linewidth]{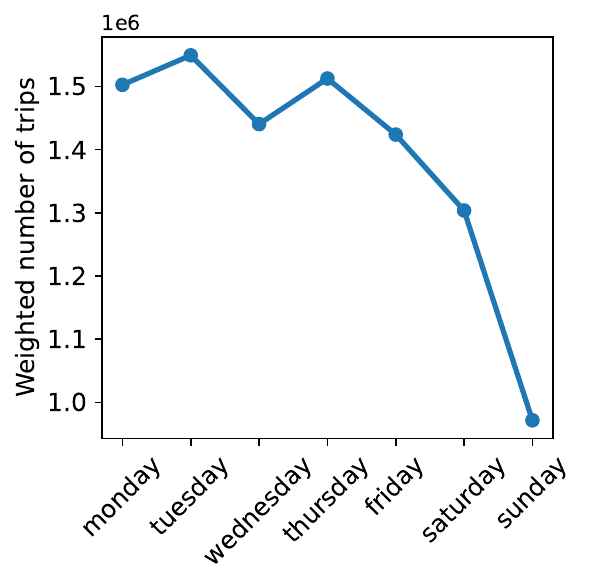}
  \caption{Walking}
\end{subfigure}
\begin{subfigure}[t]{0.32\linewidth}
  \centering
  \includegraphics[width=\linewidth]{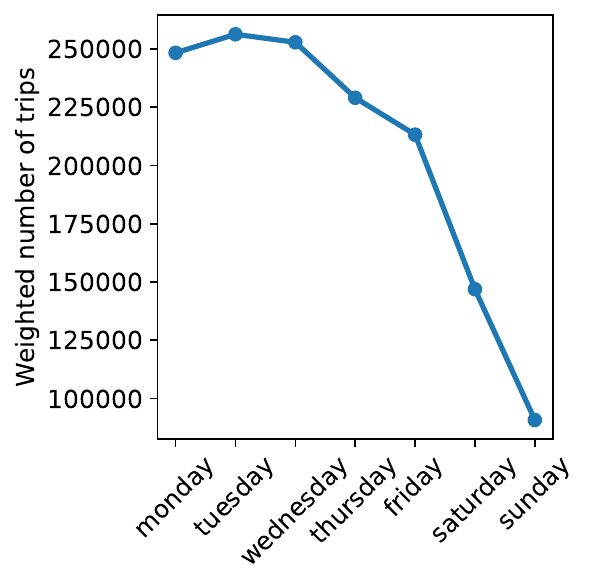}
  \caption{Bike}
\end{subfigure}

\vspace{0.5em}

\begin{subfigure}[t]{0.32\linewidth}
  \centering
  \includegraphics[width=\linewidth]{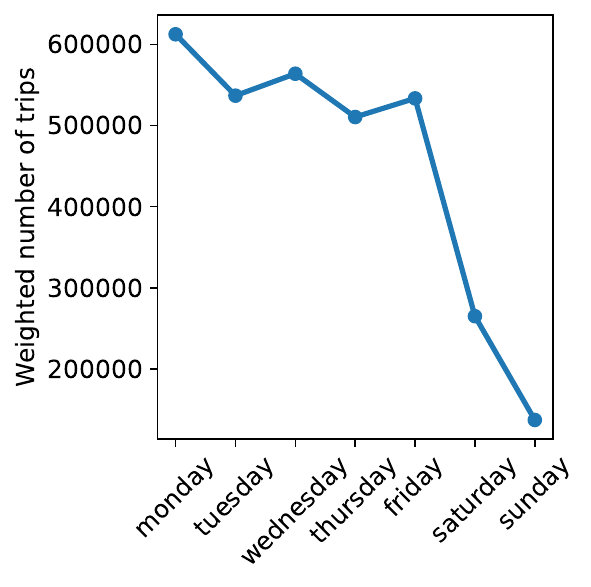}
  \caption{Train}
\end{subfigure}
\begin{subfigure}[t]{0.32\linewidth}
  \centering
  \includegraphics[width=\linewidth]{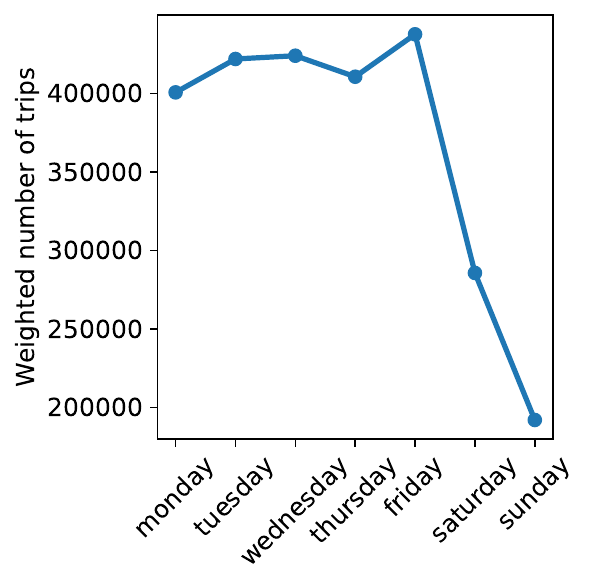}
  \caption{Subway}
\end{subfigure}
\begin{subfigure}[t]{0.32\linewidth}
  \centering
  \includegraphics[width=\linewidth]{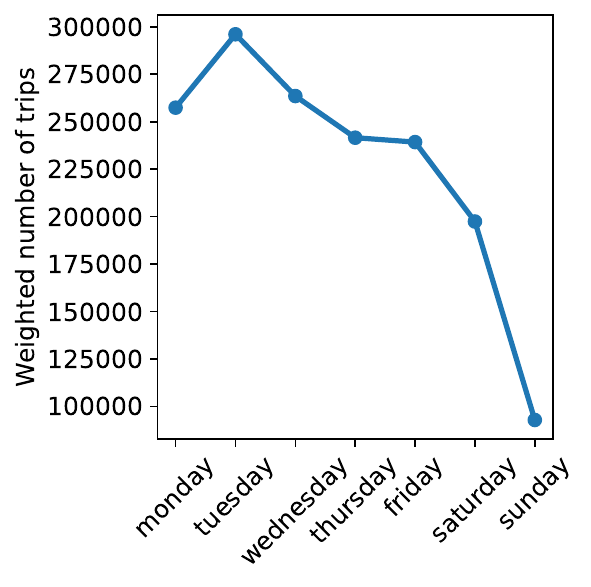}
  \caption{Bus}
\end{subfigure}

\vspace{0.5em}

\begin{subfigure}[t]{0.32\linewidth}
  \centering
  \includegraphics[width=\linewidth]{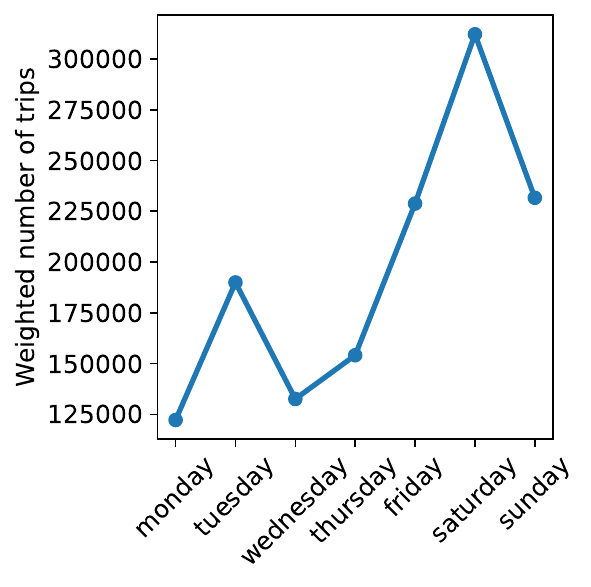}
  \caption{Private car (passenger)}
  \label{fig:passenger}
\end{subfigure}

\caption{Weighted number of trips per day of the week, by main transportation mode.}
\label{fig:mode-weekly-distribution}
\end{figure}

\paragraph{Transport mode usage across the week.}
The dataset specifies the main transportation mode for each trip via the \texttt{Main\_Mode} column, enabling a detailed examination of modal usage patterns throughout the week. As illustrated in Fig.~\ref{fig:mode-weekly-distribution}, most modes—including walking, biking, and public transit options such as bus, subway, and train—show a general decline in usage over the weekend. This trend likely reflects reduced commuting activity on non-working days. An exception is the \texttt{PASSENGER} mode (see Fig.~\ref{fig:passenger}), which exhibits increased usage during weekends, suggesting a greater prevalence of shared or family-based travel during leisure periods.

% \paragraph{Transport mode usage across the week.}
% The dataset includes the main transportation mode used for each trip, enabling the analysis of modal usage trends throughout the week. A general decrease in usage is observed over the weekend for all modes on Fig.~\ref{fig:mode-weekly-distribution}, including walking, biking, and public transport (bus, subway, and train). A notable exception is the \texttt{PASSENGER} mode (cf. Fig.  \ref{fig:passenger}), which tends to increase during the weekend, suggesting a rise in shared or family-based travel.

\begin{figure}[H]
    \centering
    
    \begin{subfigure}[t]{0.48\textwidth}
        \centering
        \resizebox{1\linewidth}{!}{%
        \begin{tikzpicture}[
            node distance=1.2cm and 2.8cm,
            scale=0.85, transform shape,
            every node/.style={font=\sffamily},
            every path/.style={->, >=latex, thick},
            main node/.style={draw, rounded corners, minimum height=0.7cm, minimum width=2cm, align=center},
            dest node/.style={draw, fill=gray!20, rounded corners, minimum height=0.6cm, minimum width=1.8cm, font=\sffamily\scriptsize},
            dashed arrow/.style={->, dashed, >=latex, thick}
            ]
            \node[main node] (BUS) {BUS};
            \node[main node, right=of BUS, yshift=2.5cm] (TRAIN) {TRAIN};
            \node[main node, right=of BUS, yshift=-2.5cm] (SUBWAY) {SUBWAY};
            \node[main node, right=of TRAIN, yshift=1.3cm] (SUBWAY2) {SUBWAY};
            \node[main node, right=of TRAIN, yshift=-1.3cm] (TRAIN2) {TRAIN};
            \node[main node, right=of SUBWAY, yshift=1.3cm] (TRAIN3) {TRAIN};
            \node[main node, right=of SUBWAY, yshift=-1.3cm] (BUS3) {BUS};
            \node[dest node, below=0.6cm of BUS] (DEST1) {Destination};
            \node[dest node, below=0.6cm of TRAIN] (DEST2) {Destination};
            \node[dest node, below=0.6cm of SUBWAY] (DEST3) {Destination};
            
            \draw (BUS) -- (TRAIN) node[midway, right] {16.3\%};
            \draw (BUS) -- (SUBWAY) node[midway, right] {4.4\%};
            \draw (TRAIN) -- (SUBWAY2) node[midway, right] {31.9\%};
            \draw (TRAIN) -- (TRAIN2) node[midway, left] {8.8\%};
            \draw (SUBWAY) -- (TRAIN3) node[midway, right] {16.8\%};
            \draw (SUBWAY) -- (BUS3) node[midway, left] {3.9\%};
            \draw[dashed arrow] (BUS) -- (DEST1) node[midway, left] {73.6\%};
            \draw[dashed arrow] (TRAIN) -- (DEST2) node[midway, left] {49.8\%};
            \draw[dashed arrow] (SUBWAY) -- (DEST3) node[midway, left] {77.0\%};
        \end{tikzpicture}
    }
        \caption{Trips starting by Bus}
    \end{subfigure}
    \hfill
    \begin{subfigure}[t]{0.48\textwidth}
        \centering
        \resizebox{1\linewidth}{!}{%
        \begin{tikzpicture}[
            node distance=1.2cm and 2.8cm,
            scale=0.85, transform shape,
            every node/.style={font=\sffamily},
            every path/.style={->, >=latex, thick},
            main node/.style={draw, rounded corners, minimum height=0.7cm, minimum width=2.3cm, align=center},
            dest node/.style={draw, fill=gray!20, rounded corners, minimum height=0.6cm, minimum width=1.8cm, font=\sffamily\scriptsize},
            dashed arrow/.style={->, dashed, >=latex, thick}
            ]
            \node[main node] (CAR) {PRIV\_CAR\_DRIVER};
            \node[main node, right=of CAR, yshift=2.2cm] (TRAIN) {TRAIN};
            \node[main node, right=of CAR, yshift=-2.2cm] (BUS) {BUS};
            \node[main node, right=of TRAIN, yshift=1.2cm] (SUBWAY2) {SUBWAY};
            \node[main node, right=of TRAIN, yshift=-1.2cm] (TRAMWAY2) {TRAMWAY};
            \node[main node, right=of BUS, yshift=1.0cm] (SUBWAY3) {SUBWAY};
            \node[main node, right=of BUS, yshift=-1.0cm] (TRAIN3) {TRAIN};
            \node[dest node, below=0.6cm of CAR] (DEST1) {Destination};
            \node[dest node, below=0.6cm of TRAIN] (DEST2) {Destination};
            \node[dest node, below=0.6cm of BUS] (DEST3) {Destination};
            
            \draw (CAR) -- (TRAIN) node[midway, right] {0.9\%};
            \draw (CAR) -- (BUS) node[midway, right] {0.1\%};
            \draw (TRAIN) -- (SUBWAY2) node[midway, right] {39.1\%};
            \draw (TRAIN) -- (TRAMWAY2) node[midway, left] {15.3\%};
            \draw (BUS) -- (SUBWAY3) node[midway, right] {3.8\%};
            \draw (BUS) -- (TRAIN3) node[midway, left] {1.4\%};
            \draw[dashed arrow] (CAR) -- (DEST1) node[midway, left] {98.7\%};
            \draw[dashed arrow] (TRAIN) -- (DEST2) node[midway, left] {34.3\%};
            \draw[dashed arrow] (BUS) -- (DEST3) node[midway, left] {94.8\%};
        \end{tikzpicture}
    }
        \caption{Trips starting by Private car}
    \end{subfigure}
    
    \vspace{1em}
    
    \begin{subfigure}[t]{0.48\textwidth}
        \centering
        \resizebox{1\linewidth}{!}{%
        \begin{tikzpicture}[
            node distance=1.2cm and 2.8cm,
            scale=0.85, transform shape,
            every node/.style={font=\sffamily},
            every path/.style={->, >=latex, thick},
            main node/.style={draw, rounded corners, minimum height=0.7cm, minimum width=2.3cm, align=center},
            dest node/.style={draw, fill=gray!20, rounded corners, minimum height=0.6cm, minimum width=1.8cm, font=\sffamily\scriptsize},
            dashed arrow/.style={->, dashed, >=latex, thick}
            ]
            \node[main node] (TRAIN) {TRAIN};
            \node[main node, right=of TRAIN, yshift=2.5cm] (SUBWAY) {SUBWAY};
            \node[main node, right=of TRAIN, yshift=-2.5cm] (BUS) {BUS};
            \node[main node, right=of SUBWAY, yshift=1.3cm] (SUBWAY2) {SUBWAY};
            \node[main node, right=of SUBWAY, yshift=-1.3cm] (BUS2) {BUS};
            \node[main node, right=of BUS, yshift=1.1cm] (TRAIN2) {TRAIN};
            \node[main node, right=of BUS, yshift=-1.1cm] (BUS3) {BUS};
            \node[dest node, below=0.6cm of TRAIN] (DEST1) {Destination};
            \node[dest node, below=0.6cm of SUBWAY] (DEST2) {Destination};
            \node[dest node, below=0.6cm of BUS] (DEST3) {Destination};
            
            \draw (TRAIN) -- (SUBWAY) node[midway, right] {27.5\%};
            \draw (TRAIN) -- (BUS) node[midway, right] {11.4\%};
            \draw (SUBWAY) -- (SUBWAY2) node[midway, right] {7.7\%};
            \draw (SUBWAY) -- (BUS2) node[midway, left] {4.9\%};
            \draw (BUS) -- (TRAIN2) node[midway, right] {2.8\%};
            \draw (BUS) -- (BUS3) node[midway, left] {1.4\%};
            \draw[dashed arrow] (TRAIN) -- (DEST1) node[midway, left] {45.2\%};
            \draw[dashed arrow] (SUBWAY) -- (DEST2) node[midway, left] {82.5\%};
            \draw[dashed arrow] (BUS) -- (DEST3) node[midway, left] {93.8\%};
        \end{tikzpicture}
    }
        \caption{Trips starting by Train}
    \end{subfigure}
    \hfill
    \begin{subfigure}[t]{0.48\textwidth}
        \centering
        \resizebox{1\linewidth}{!}{%
        \begin{tikzpicture}[
            node distance=1.2cm and 2.8cm,
            scale=0.85, transform shape,
            every node/.style={font=\sffamily},
            every path/.style={->, >=latex, thick},
            main node/.style={draw, rounded corners, minimum height=0.7cm, minimum width=2.3cm, align=center},
            dest node/.style={draw, fill=gray!20, rounded corners, minimum height=0.6cm, minimum width=1.8cm, font=\sffamily\scriptsize},
            dashed arrow/.style={->, dashed, >=latex, thick}
            ]
            \node[main node] (SUBWAY) {SUBWAY};
            \node[main node, right=of SUBWAY, yshift=2.5cm] (TRAIN) {TRAIN};
            \node[main node, right=of SUBWAY, yshift=-2.5cm] (SUBWAY2) {SUBWAY};
            \node[main node, right=of TRAIN, yshift=1.3cm] (BUS) {BUS};
            \node[main node, right=of TRAIN, yshift=-1.3cm] (CAR) {PRIV\_CAR\_DRIVER};
            \node[main node, right=of SUBWAY2, yshift=1.3cm] (TRAIN2) {TRAIN};
            \node[main node, right=of SUBWAY2, yshift=-1.3cm] (SUBWAY3) {SUBWAY};
            \node[dest node, below=0.6cm of SUBWAY] (DEST1) {Destination};
            \node[dest node, below=0.6cm of TRAIN] (DEST2) {Destination};
            \node[dest node, below=0.6cm of SUBWAY2] (DEST3) {Destination};
            
            \draw (SUBWAY) -- (TRAIN) node[midway, right] {17.4\%};
            \draw (SUBWAY) -- (SUBWAY2) node[midway, right] {10.8\%};
            \draw (TRAIN) -- (BUS) node[midway, right] {19.1\%};
            \draw (TRAIN) -- (CAR) node[midway, left] {5.1\%};
            \draw (SUBWAY2) -- (TRAIN2) node[midway, right] {8.8\%};
            \draw (SUBWAY2) -- (SUBWAY3) node[midway, left] {7.1\%};
            \draw[dashed arrow] (SUBWAY) -- (DEST1) node[midway, left] {64.0\%};
            \draw[dashed arrow] (TRAIN) -- (DEST2) node[midway, left] {68.8\%};
            \draw[dashed arrow] (SUBWAY2) -- (DEST3) node[midway, left] {81.1\%};
        \end{tikzpicture}
    }
        \caption{Trips starting by Subway}
        
    \end{subfigure}
    \vspace{1em}
    
    \begin{subfigure}[t]{0.48\textwidth}
        \centering
        \resizebox{1\linewidth}{!}{%
        \begin{tikzpicture}[
            node distance=1.2cm and 3.5cm,
            every node/.style={font=\sffamily},
            every path/.style={->, >=latex, thick},
            main node/.style={draw, rounded corners, minimum height=0.7cm, minimum width=2.6cm, align=center},
            dest node/.style={draw, fill=gray!20, rounded corners, minimum height=0.6cm, minimum width=2.2cm, font=\sffamily\scriptsize},
            dashed arrow/.style={->, dashed, >=latex, thick}
            ]
            
            % Niveau 0
            \node[main node] (BIKE) {BIKE};
            
            % Niveau 1 : Mode_2
            \node[main node, right=of BIKE, yshift=2.4cm] (TRAIN) {TRAIN};
            \node[main node, right=of BIKE, yshift=-2.4cm] (SUBWAY) {SUBWAY};
            
            % Niveau 2 : Mode_3 pour TRAIN
            \node[main node, right=of TRAIN, yshift=1.5cm] (BIKE2) {BIKE};
            \node[main node, right=of TRAIN, yshift=-1.5cm] (BUS2) {BUS};
            
            % Niveau 2 : Mode_3 pour SUBWAY
            \node[main node, right=of SUBWAY, yshift=1.3cm] (BUS3) {BUS};
            \node[main node, right=of SUBWAY, yshift=-1.3cm] (SUBWAY2) {SUBWAY};
            
            % Destinations
            \node[dest node, below=0.6cm of BIKE] (DEST1) {Destination};
            \node[dest node, below=0.6cm of TRAIN] (DEST2) {Destination};
            \node[dest node, below=0.6cm of SUBWAY] (DEST3) {Destination};
            
            % Arêtes BIKE → Mode_2
            \draw (BIKE) -- (TRAIN) node[midway, right] {4.7\%};
            \draw (BIKE) -- (SUBWAY) node[midway, right] {0.7\%};
            
            % Arêtes TRAIN → Mode_3
            \draw (TRAIN) -- (BIKE2) node[midway, right] {20.7\%};
            \draw (TRAIN) -- (BUS2) node[midway, left] {8.3\%};
            
            % Arêtes SUBWAY → Mode_3
            \draw (SUBWAY) -- (BUS3) node[midway, right] {36.6\%};
            \draw (SUBWAY) -- (SUBWAY2) node[midway, left] {16.5\%};
            
            % Flèches vers Destination
            \draw[dashed arrow] (BIKE) -- (DEST1) node[midway, left] {94.2\%};
            \draw[dashed arrow] (TRAIN) -- (DEST2) node[midway, left] {57.4\%};
            \draw[dashed arrow] (SUBWAY) -- (DEST3) node[midway, left] {45.3\%};
            
        \end{tikzpicture}
    }
    \caption{Weighted travel chains for trips starting by \texttt{BIKE}.}
    \end{subfigure}
    \caption{Weighted travel chains starting from four main transportation modes.}
    \label{fig:graphes_modes}
 \end{figure}
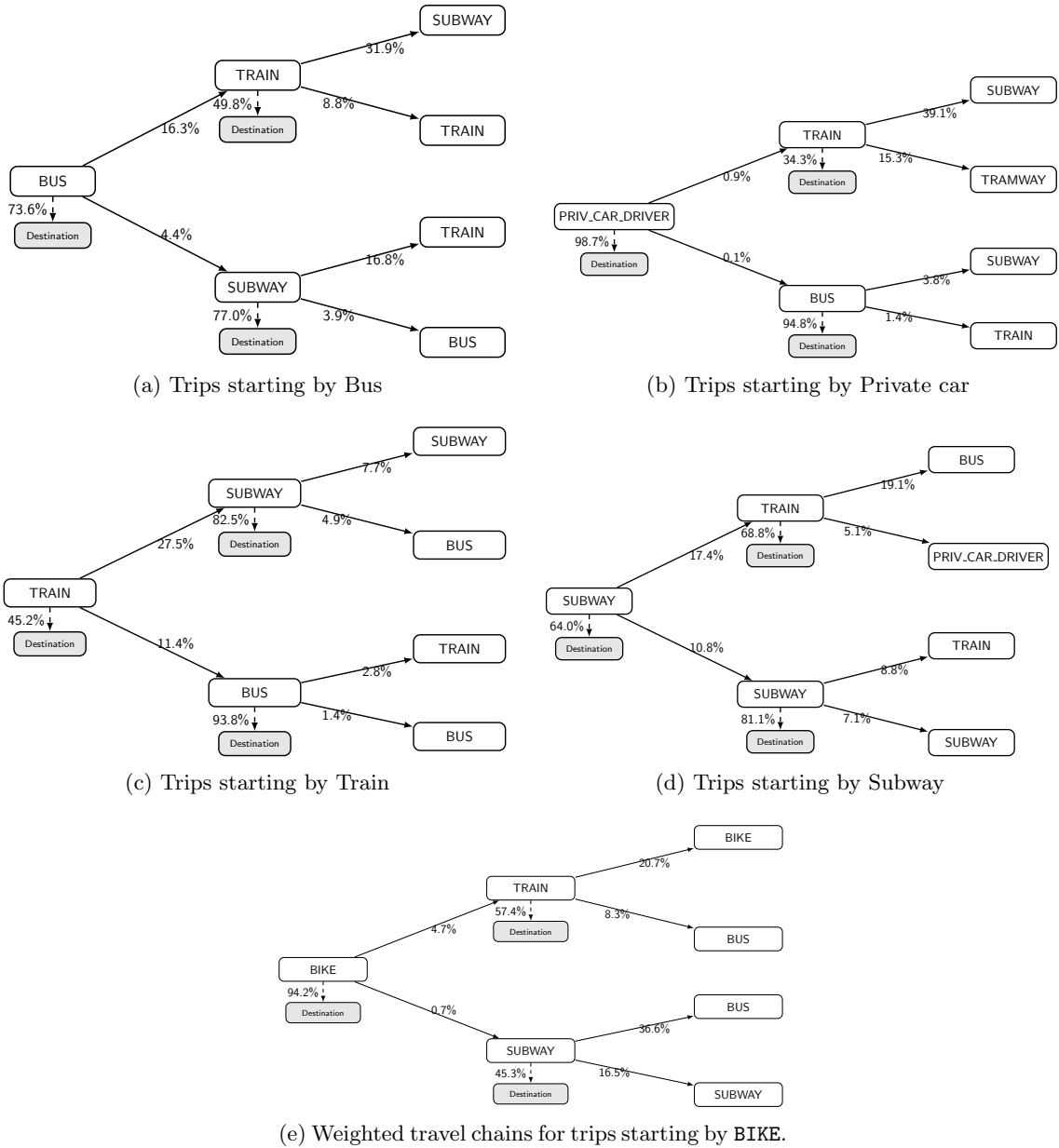

\paragraph{Multimodal trip analysis.}
The dataset includes detailed transport sequences, enabling the reconstruction of multimodal trip chains. For each main transport mode—such as \texttt{BUS}, \texttt{TRAIN}, \texttt{SUBWAY}, \texttt{PRIV\_CAR\_DRIVER}, and \texttt{BIKE}—we identify the most frequent subsequent modes and the proportion of trips that terminate at each stage. These patterns are illustrated in Fig.~\ref{fig:graphes_modes}.

The resulting visualizations reveal the structural characteristics of multimodal mobility. For instance, trips starting with \texttt{BUS} often continue with either \texttt{SUBWAY} or \texttt{TRAIN}, indicating integration with regional transit networks. In contrast, trips initiated by \texttt{BIKE} typically lead directly to the final destination. Likewise, trips that begin with \texttt{WALKING} almost always conclude without transitioning to another mode, underscoring its role in short-distance or final-leg travel.

\begin{figure}[!h]
\centering
\begin{subfigure}{0.48\linewidth}
\includegraphics[width=\linewidth]{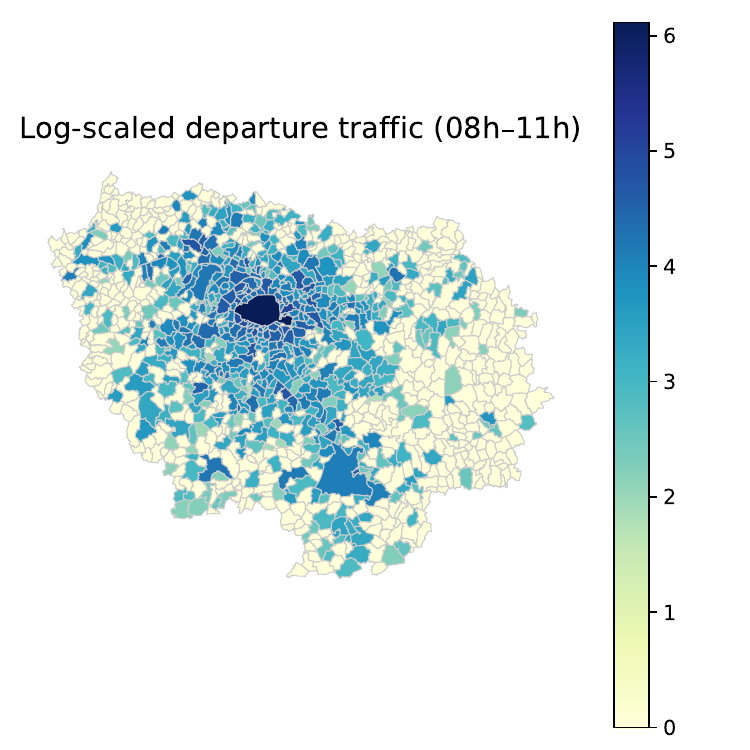}
\caption{08h–11h}
\end{subfigure}
\begin{subfigure}{0.48\linewidth}
\includegraphics[width=\linewidth]{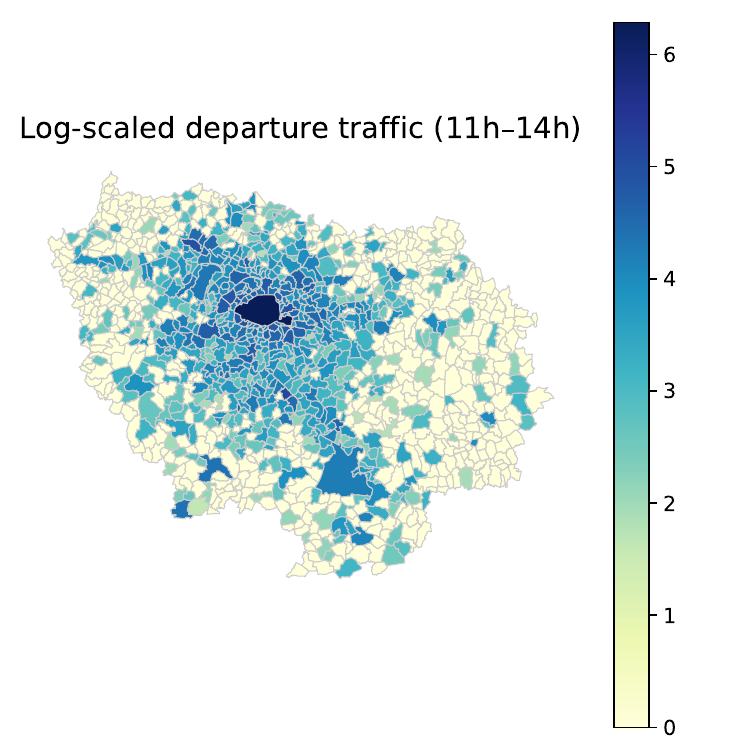}
\caption{11h–14h}
\end{subfigure}

\begin{subfigure}{0.48\linewidth}
\includegraphics[width=\linewidth]{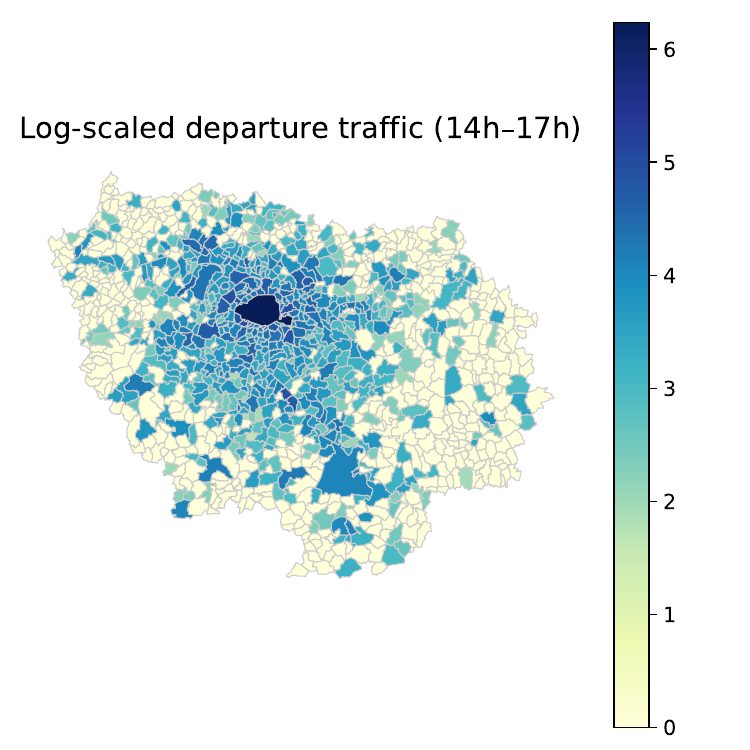}
\caption{14h–17h}
\end{subfigure}
\begin{subfigure}{0.48\linewidth}
\includegraphics[width=\linewidth]{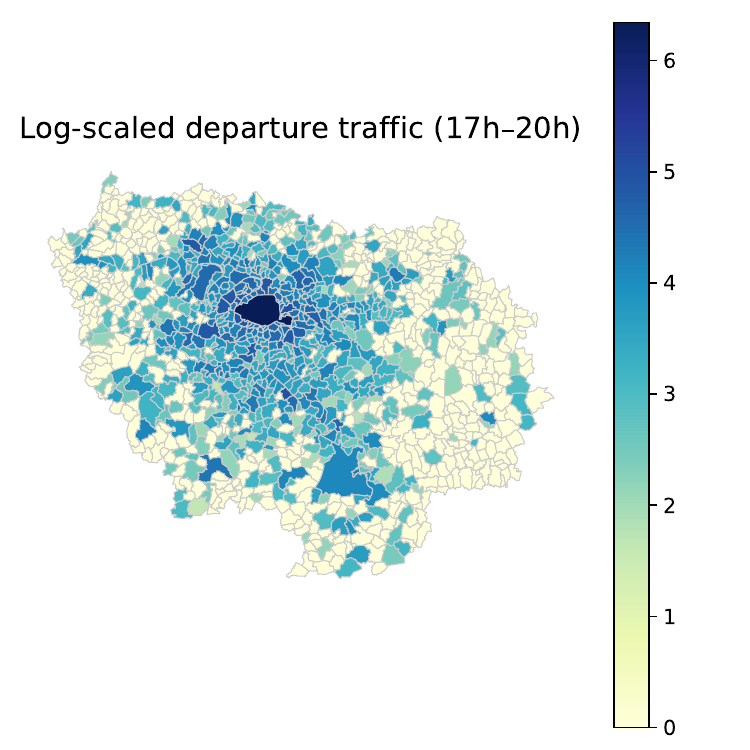}
\caption{17h–20h}
\end{subfigure}
\caption{Log-scaled spatial distribution of departure locations throughout the day.}
\label{fig:departure-map}
\end{figure}

% \paragraph{Departure flow statistics.}
% By aggregating trip origins per city (\texttt{Area\_O}) and weighting by individual trip weights, we build in Fig. \ref{fig:departure-map}, a map showing the spatial distribution of departures over the day. Given the overrepresentation of large municipalities (e.g., Paris), the maps use a logarithmic color scale to enhance the readability of smaller areas and support regional comparisons.
\paragraph{Departure flow statistics.}
By aggregating trip origins per municipality (column \texttt{Area\_O}) and weighting by individual trip weights (\texttt{Weight\_Day}), we generate the spatial distribution of departures across the Île-de-France region, as shown in Fig.~\ref{fig:departure-map}. To improve visual interpretability and enable meaningful regional comparisons, particularly in the presence of dominant urban centers such as Paris, the map employs a logarithmic color scale. This representation enhances the visibility of departure patterns in less populated municipalities that would otherwise be overshadowed.

\begin{figure}[!h]
\centering
\begin{subfigure}{0.48\linewidth}
\includegraphics[width=\linewidth]{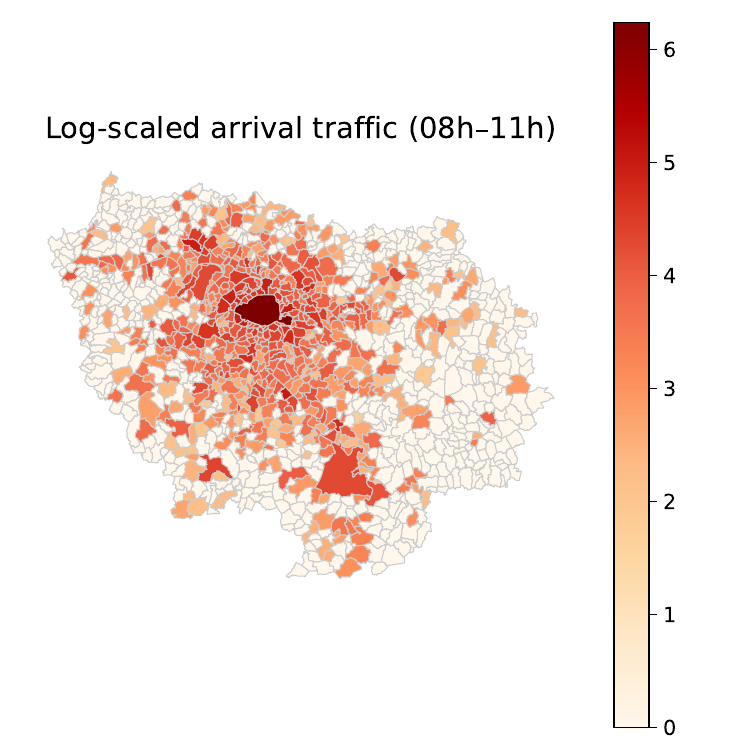}
\caption{08h–11h}
\end{subfigure}
\begin{subfigure}{0.48\linewidth}
\includegraphics[width=\linewidth]{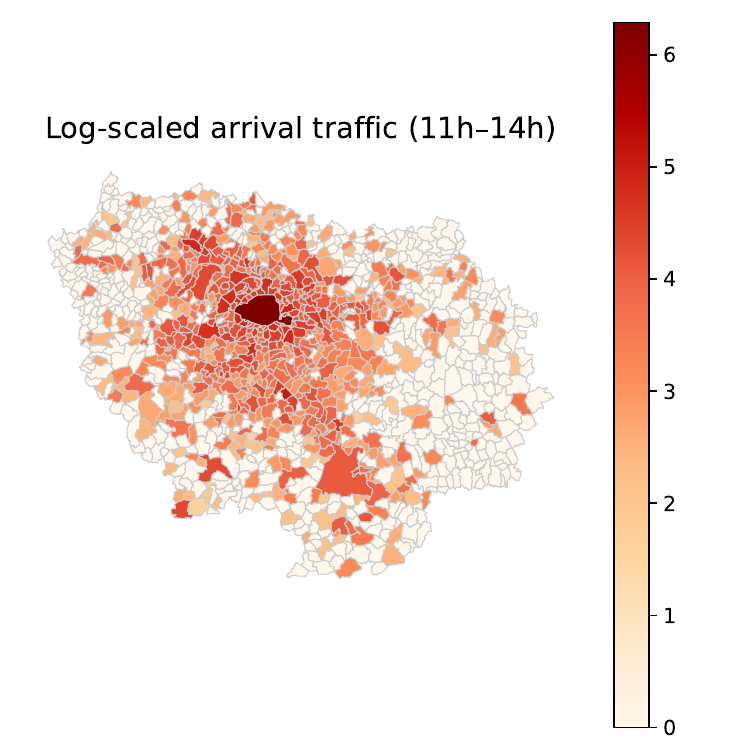}
\caption{11h–14h}
\end{subfigure}

\begin{subfigure}{0.48\linewidth}
\includegraphics[width=\linewidth]{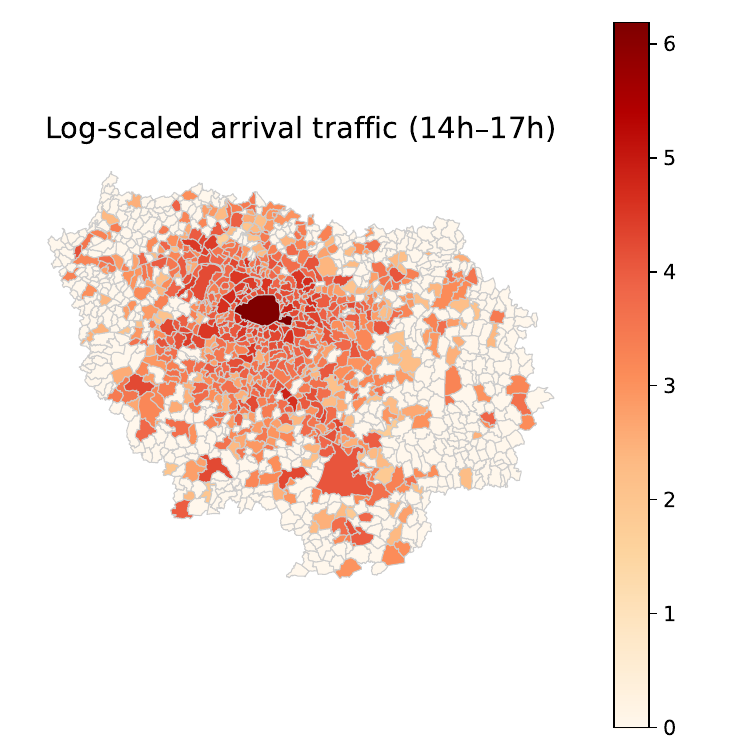}
\caption{14h–17h}
\end{subfigure}
\begin{subfigure}{0.48\linewidth}
\includegraphics[width=\linewidth]{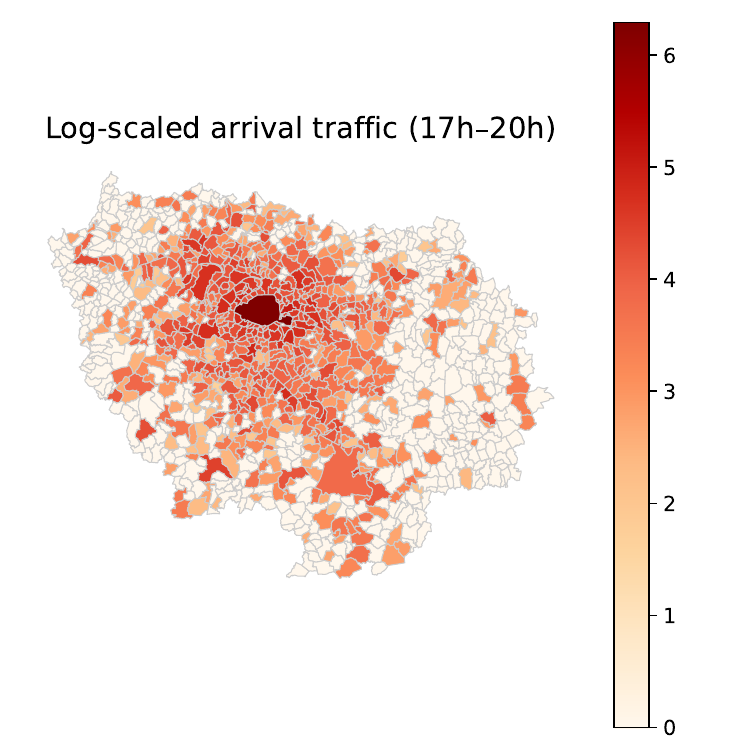}
\caption{17h–20h}
\end{subfigure}
\caption{Log-scaled spatial distribution of arrival locations throughout the day.}
\label{fig:arrival-map}
\end{figure}

\paragraph{Arrival flow statistics.}
A similar methodology is applied to trip destinations, based on the \texttt{Area\_D} column, resulting in the spatial distribution of arrivals shown in Fig.~\ref{fig:arrival-map}. As with departures, a logarithmic scale is used to mitigate the visual dominance of major cities. Despite this normalization, Paris and the inner suburbs remain disproportionately represented in both departure and arrival volumes, underscoring the central role these areas play in regional mobility dynamics and highlighting the core–periphery structure of commuting flows in Île-de-France.
 
% \section{User Behavior on GPS Traces}
% Records per user/day/hour

% Unique locations per user

% Inter-event time

% Days with valid tracking

\section{Methods for Future Analyses}
This section outlines possible analytical directions enabled by the dataset, along with practical guidelines for appropriate usage—particularly with respect to weighting and spatial granularity.

\paragraph{Analytical potential.}  
The richness and scale of the dataset open doors to a wide array of studies in transportation modeling, mobility behavior, infrastructure planning, and equity assessments. Examples include:
\begin{itemize}[leftmargin=*]
    \item Estimating modal shift under various policy scenarios;
    \item Modeling accessibility and travel time disparities;
    \item Investigating urban-rural differences in multimodal behavior;
    \item Studying travel demand at different temporal and spatial resolutions.
\end{itemize}

\paragraph{Use of weights.}  
To draw population-level conclusions, it is essential to apply the provided individual and trip-level weights appropriately. These weights correct for sampling biases across sociodemographic categories, time, and geography. Researchers should always consider whether to normalize weights, especially in comparative or regression-based analyses, and whether to disaggregate by day type or geographic unit.

\paragraph{Spatial data.}  
The dataset can be analyzed at multiple spatial scales. Complementary shapefiles are provided in the \href{https://gitlab.inria.fr/netmob2025/data-challenge/-/tree/main/data}{\texttt{data/}} directory of the  GitLab repository:
\begin{itemize}
    \item \texttt{communes\_france.geojson}: Polygons of all French communes;
    \item \texttt{paris\_iris.geojson}: IRIS-level subdivisions within Paris;
    \item \texttt{paris\_visited\_iris.geojson}: IRIS polygons actually visited in the dataset.
\end{itemize}
These files allow fine-grained geographic aggregation, particularly in Paris and its surrounding communes.

\paragraph{Further documentation.}  
Additional technical details and usage guidance can be found in the official dataset repository. The directory \href{https://gitlab.inria.fr/netmob2025/data-challenge/-/tree/main/docs}{\texttt{docs/}} includes slide decks and technical notes on the data.

\section{Conclusion and Availability}

This document has presented the structure, processing pipeline, and potential applications of the \netmob dataset, a large-scale, survey-augmented mobility dataset collected in Île-de-France. We detailed the content of the individual, trip, and GPS trajectory databases, the anonymization and spatial filtering steps, and the methodology for deriving and applying weights to ensure representativity across the region’s population.

\paragraph{Dataset access.}  
The dataset is accessible in the framework of the \netmob data challenge, subject to acceptance of the terms and conditions available at the following link:\footnote{\url{https://drive.google.com/file/d/13Jssi358EDWu9Jf4DROTWMiPwVZwglzy/view}}. Upon approval, participants can access the full dataset and associated documentation through the official GitLab repository:\footnote{\url{https://gitlab.inria.fr/netmob2025/data-challenge}}.

\paragraph{Intended use and impact.}  
The \netmob dataset is intended to support a wide range of studies in urban mobility, transportation equity, and behavioral modeling. By combining detailed GPS trajectories with rich sociodemographic annotations and calibrated statistical weights, it offers a valuable resource for researchers and practitioners aiming to understand and improve mobility systems across demographic and spatial dimensions. We hope this dataset contributes to inclusive, data-driven transport policy and advances in mobility data science.

{\small
\bibliographystyle{plain}
\bibliography{references}
}

\end{document}